\documentclass{aa}  

\usepackage{graphicx}
\usepackage{txfonts}
\usepackage{lipsum}
\usepackage{subcaption}         
\usepackage{lscape}             
\usepackage{placeins}           
                                
\begin{document}

   \title{MHD Modelling of magnetic reconnection heating and jets in the solar corona}

%
%
%

\author{P. Pagano\inst{1,2}
          \and
          C. Argiroffi\inst{1,2}
          \and
          F. Reale\inst{1,2}
          \and
          A. Petralia\inst{2}
          \and
          G. Cozzo\inst{3}
          \and
          P. Testa\inst{3}          }

   \institute{
            Dipartimento di Fisica \& Chimica, Università di Palermo, Piazza del Parlamento 1, I-90134 Palermo, 
            \and
            INAF-Osservatorio Astronomico di Palermo, Piazza del Parlamento 1, I-90134 Palermo, Italy
            \and
            Harvard–Smithsonian Center for Astrophysics, 60 Garden St., Cambridge, MA 02193, USA
            \email{paolo.pagano@unipa.it}
            }
   \date{Received \dots ; accepted \dots}

 
  \abstract
   {Magnetic reconnection is seen as the lead mechanism to explain the persistent and bursty heating of the solar corona that keeps its thermal structure. Lately the observations and modelling of magnetic reconnection outflows, i.e. nanojets, has empowered our capabilities to scrutinise this mechanisms, however more needs to be understood on the relation between the energetic and kinematic events.}
   {The aim of this paper is to study in details the properties of the heating events (nanoflare) and the kinematic counterparts (nanojets) in order to augment detection capabilities of magnetic reconnection events.}
   {We use MHD simulations of the interaction of two magnetic flux tubes where magnetic reconnection occurs and such reconnection outflows are generated. We vary key parameters, such as the critical current and the size of the domain to study the heating and slingshot mechanisms.}
   {We find that the properties of the nanojets can give away much information on the magnetic reconnection that generated it, while it is difficult to directly link the amount of heating and the jet speeds. We also find that such reconnection model can explain a wide range of heating events.}
   {Magnetic reconnection still is invariably linked to both heating and outflows and our simulations show the complex interplay between these two byproducts of the magnetic reconnection and how their properties can be studied in observations.}


   \maketitle
   \nolinenumbers
%

\section{Introduction} \label{sec:introduction}

Magnetic reconnection is scrutinised with increasing interest as the key mechanism behind the persistent, bursty conversion of magnetic energy into thermal energy in the solar corona.
This is the well known coronal heating problem: we have long diagnosed the multimillion-degree thermal structure of the solar corona, and identified the magnetic field as the obvious energy supply to maintain it against radiative losses and thermal conduction. However, the mechanism converting magnetic into thermal energy remains elusive and hard to observe, both because of its impulsive, short-lived occurrence and the difficulty of disentangling magnetic structures evolving in an optically thin corona \citep{ParnellDeMoortel2012, Klimchuk2015, DeMoortelBrowning2015}

Despite a number of models successfully accounting for plasma heating from different magnetic processes \citep{2020SSRv..216..140V, 2022FrASS...920116A, 2023ApJ...957...25J}, this paper focuses on features of magnetic reconnection, a process at the foundation of the nanoflares model \citep{Parker1988}, where a myriad of such events maintain the plasma temperature through their collective contribution.
This theory has always attracted interest, as it offers several features in common with observations.
This theory envisages a bursty, intermittent, and localised heating, which are all the key features that decades of observations have helped identifying for the coronal heating mechanism \citep{Klimchuk2015}.

Coronal heating is best studied in coronal loops \citep{2014LRSP...11....4R}, as they are relatively long-lived, stable structures where intense magnetic fields have clear configurations, making models easier to connect to observations. 
Moreover, coronal loops, especially in active regions where they interact with several other magnetic flux tubes, are the ideal environment for the slow braiding of magnetic field lines — the precursor of the sudden magnetic reconnection that converts all at once the energy slowly accumulated during braiding.
\citet{2015RSPTA.37340260C} offers one of the most thorough analyses linking observations to nanoflares.
At the same time, it is increasingly common to investigate heating in solar filaments and prominences as interesting sites for studying plasma heating \cite{2014LRSP...11....1P, 2022NatAs...6..942J, 2025ApJ...985L..12G}.
Additionally, \citet{2025ApJ...995L..65R} have shown a strong link between magnetic field intensity and the heating available in coronal loops.

This nanoflare scenario has come under even closer scrutiny since \citet{2021NatAs...5...54A} showed that it is in principle possible to detect its dynamical counterpart, the nanojets — magnetic reconnection outflows springing from the reconnection region because of the slingshot effect. 
These observations and modelling proved that we can measure not only the cumulative effect of magnetic reconnection events but sometimes also single ones, making it worth investigating their physical properties.
Since then, several works have analysed single magnetic reconnection events, and the literature on such reconnection outflows, i.e. nanojets, has expanded, with nanojets observed in many situations and with a number of instruments
\citep{2022ApJ...938..122P, 2022ApJ...934..190S, 2024A&A...692A.119C,2025ApJ...985L..12G,2025ApJ...982..147M, 2025ApJ...988L..65B, 2025A&A...702A.189T, 2025ApJ...995...94C, 2025MNRAS.544.1758W}.
Simultaneously, more modelling efforts have focused on understanding nanojet dynamics, as well as synthesising observables to sharpen our capability to detect them and study their connection with coronal heating \citep{2021A&A...656A.141P, 2022ApJ...926...52D, 2023A&A...678A..40C, 2023MNRAS.518.1584R, 2024A&A...689A.184C, 2024MNRAS.530.2361B, 2025A&A...695A..40C, 2025A&A...699A.106S, 2026ApJ...998...75C, 2026ApJ...998...76C}.
More generally, \citet{2015RSPTA.37340265W} and \citet{2020LRSP...17....5P} provide elaborate summaries of the models developed to explain how magnetic braiding and reconnection can account for coronal heating.

A number of works have recently shown that braiding of magnetic field lines from the photosphere can self-consistently provide sufficient energy flux to support the corona's thermal structure \citep{2005ApJ...618.1020G, 2005ApJ...618.1031G, 2011A&A...530A.112B, 2015ApJ...811..106H, 2017ApJ...834...10R, 2019A&A...624L..12W, 2026ApJ..1004..149C}. This work thus focuses on the mechanisms that follow magnetic reconnection, to shed light on the last segment of the conversion of free magnetic energy into heating and kinetic energy.

This work aims to build more knowledge on the mechanisms that generate nanoflares and how they evolve, with multiple specific aims.
First, we want to identify features that favour magnetic braiding leading to magnetic reconnection.
Magnetic reconnection naturally occurs when boundary motions braid the coronal magnetic field \citep{1988GApFD..41..181V, 1989ApJ...338.1148M, 2015ApJ...805...47P}, and we aim to illustrate how the mechanism can progress against the restoring reaction of the pressure forces that inevitably build up at the current sheet.

Moreover, we want to investigate possible links between nanoflares (i.e., the heating) and nanojets (i.e., the velocity), or whether the properties of one can become a proxy for the properties of the other. For example, is there a correlation between the heating or maximum temperature at the reconnection site and the velocity of the jets? This is a crucial question now, as the way is being paved towards detecting more nanojets, but we still want to benchmark these events against the effects of the myriad, still elusive, nanoflares. 
Still on the issue of theoretical confirmation for the nanoflare theory, we intend to run MHD simulations to prove that, at the correct physical scale, magnetic reconnection can account for single events across different energetics, including those predicted by the nanoflare theory.

Finally, we perform a simple but effective forward modelling of the $284.16~\AA$ and $108.36~\AA$ emission lines from Fe XV and Fe XIX ions, respectively. These lines probe different thermal conditions of coronal plasma (2.5 MK and 10 MK, respectively) and are among those selected for the forthcoming Multi-slit Solar Explorer (MUSE) mission \citep{2020ApJ...888....3D}.

To pursue these aims we resort to MHD simulations, modelling the evolution of magnetic flux tubes resembling coronal loops. In our model, the footpoint motion first induces a slow evolution and then triggers magnetic reconnection as a consequence of the magnetic configuration it attains.
The model builds on the work of \citet{2014A&A...564A..48G}, \citet{Reale2016}, and \citet{2021NatAs...5...54A}.

The paper is structured as follows: in Sec.\ref{sec:model} we describe the model,
in Sec.\ref{sec:results} we illustrate the results of the MHD simulations and the exploration of their parameter space,
in Sec.\ref{sec:lines} we describe and illustrate the method and results of the synthesis of Fe XV and Fe XIX lines,
and finally in Sec.\ref{sec:conclusions} we discuss our results and draw some conclusions.

\section{The model} \label{sec:model}

We use a model of two parallel flux tubes extending in a 3D cartesian domain. The flux tubes represent a pair of coronal loops or loop strands that connect two locations in the chromosphere. Thus, the vertical direction is used for the longitudinal direction along the guide magnetic field of the loops and the two horizontal directions are the two transverse directions perpendicular to the guide field.
The chromosphere is treated as a nearly infinite mass reservoir for the solar corona that is placed in the center of our cartesian domain between the two chromospheres.

This model with two flux tubes has already been presented in \citet{2021NatAs...5...54A} and is built on the single loop model of \citet{Reale2016}. More recently, similar setups stemmed from this one, as in \citet{2023Symm...15..627C, 2023A&A...678A..40C, 2024A&A...689A.184C, 2025A&A...695A..40C, 2026ApJ...998...75C, 2026ApJ...998...76C}.
The initial conditions are constructed from the relaxation of a 1D initial guess of a coronal loop density and temperature distribution (i.e. along the guide field) which is then exported in a 3D cartesian domain where a magnetic flux concentration is present and persists at the chromospheric boundaries.
Such a system is let relax numerically by solving the Magnetohydrodynamics (MHD) equations, with the PLUTO code \citep{Mignone_2012ApJS..198....7M}.
The equations are solved in Eulerian, conservative form:
\begin{equation}
\frac{\partial\rho}{\partial t} + \mathbf{\nabla} \cdot (\rho\vec{v})= 0,
\label{eq:mass_conservation}
\end{equation}
\begin{equation}
\frac{(\partial \rho \vec{v})}{\partial t} + \mathbf{\nabla} \cdot (\rho \vec{v} \vec{v}) = - \nabla \cdot \left(P\mathbf{I} + \frac{B^2}{8 \pi}\mathbf{I} - \frac{\vec{B}\vec{B}}{8\ \pi}\right)+ \rho\vec{g},
\label{eq:momentum} \\
\end{equation}
\begin{equation}
\frac{\partial \vec{B}}{\partial t} - \nabla \times \left(\vec{v}\times\vec{B}\right)=  \eta \nabla^2 \vec{B},
\label{induction_eq} \\
\end{equation}
\begin{equation}
\begin{split}
\frac{\partial}{\partial t} \left(\frac{B^2}{8 \pi} + \frac{1}{2} \rho v^2 + \rho \epsilon\right)
+ \nabla \cdot \left[ \frac{c}{4 \pi} \vec{E} \times \vec{B}+ \frac{1}{2}\rho v^2 \vec{v}+ \frac{\gamma}{\gamma-1} P\vec{v}
+ \vec{F_c}\right]\\
\quad =- \Lambda(T) n_e n_H + H_0
\end{split}
\label{energy_eq} 
\end{equation} 
\begin{equation}
P = (\gamma -1) \rho \epsilon = \frac{2 k_B}{\mu m_H} \rho T,
\label{eq: EOS}
\end{equation}
\begin{equation}
\vec{j}= \frac{c}{4 \pi} \mathbf{\nabla} \times \vec{B}
\label{Eq:current_density}
\end{equation}
\begin{equation}
\vec{E}= -  \frac{\vec{v}}{c} \times \vec{B} + \frac{\vec{j}}{\sigma},
\label{Eq:electric_field}
\end{equation}
where $t$ is the time; 
$\rho$ the mass density; 
$\vec{v}$ the plasma velocity; $P$ the thermal pressure; 
$\vec{B}$ the magnetic field; $\vec E$ the electric field;
$\vec{g}$ 
the gravity acceleration vector for a curved loop;  $\mathbf{I}$ the
identity tensor;
$\epsilon$ the internal energy; 
$\vec{j}$ the induced current density; 
$\eta$ the magnetic diffusivity;  
$\sigma = \frac{c^2}{4 \pi \eta}$ the electrical conductivity;
$T$ the temperature; 
$\vec{F}_c$ the thermal conductive flux; $\Lambda (T)$  the optically thin radiative losses per unit emission measure; $n_H$ and $n_e$  the hydrogen and electron number density, respectively; $m_H$ the hydrogen mass density; $k_B$  the Boltzmann constant;
$\mu = 1.265$ the mean ionic weight \citep[relative to a proton and assuming metal abundance of solar values: $X\,(\mathrm{H}) \simeq 70.7\,\%$, $Y\,(\mathrm{He}) \simeq 27.4\,\%$, $Z\,(\mathrm{Li-U}) \simeq 1.9\,\%$;][]{1989GeCoA..53..197A}; 
and $H_0 = 4.2 \times 10^{-5}\,\mathrm{erg}\,\mathrm{cm}^{-3}\,\mathrm{s}^{-1}$ a volumetric heating rate which balances the initial energy losses and is used to keep the loop initially in thermal equilibrium, and which is guessed for this model from the Rosner, Tucker, Vaiana (RTV) scaling laws \citep{Rosner1978}.
As shown in Eq. (\ref{eq: EOS}), we use the ideal gas law as an equation of state. 

The flux tube is circularly curved only in the corona, and that it should stay straight vertical in the chromosphere, we consider the gravity of a curved loop in the corona \citep[as already implemented in][]{Reale2016}.
The thermal conductive flux is split into its components along and across the magnetic field as in \citet{Reale2016}.
The optically thin radiative losses per unit emission measure are parametrised in PLUTO, being derived from the CHIANTI v.~7.0 database \citep{2013ApJ...772...71L}, assuming coronal element abundances from \citep{1992PhyS...46..202F}.

The model also needs to account for the steep number density and temperature change, by two orders of magnitude across the transition region in less than $100\,\mathrm{km}$. Numerically resolving such a transition is computationally challenging and not relevant here, so we adopt the Linker--Lionello--Mikić method \citep{2001JGR...10625165L,2009ApJ...690..902L,2013ApJ...773...94M}, which artificially broadens the transition region. In our simulation we set a temperature threshold $T_c = 2.5 \times 10^5\,\mathrm{K}$ for this method. 
Finally, we derive a volumetric heating rate $H_0$ sufficient to keep the corona static with an apex temperature of about $8 \times 10^{5}\,\mathrm{K}$ and half length $L = 2.5 \times 10^9\,\mathrm{cm}$, from the Rosner-Tucker-Vaiana scaling laws \citep{Rosner1978}.
In this way, we start from a background atmosphere in agreement with the hydrostatic loop model by \cite{1981ApJ...243..288S} and \cite{2014A&A...564A..48G}.

\subsection{The plasma resistivity}

We consider an anomalous plasma resistivity that is switched on only in the corona and transition region (i.e., above $T_{\mathrm{cr.}} = 10^4\,\mathrm{K}$) where the magnitude of the current density exceeds a critical value, as in the following equation \citep[e.g.][]{2009A&A...506..913H}:
\begin{equation}
    \eta =
    \begin{cases}
    \eta_0 & |J| \ge j_{\mathrm{cr}}  \text{ and } T \ge T_{\mathrm{cr.}} \\
    0  & |J| < j_{\mathrm{cr}} \text{ or } T < T_{\mathrm{cr}}
    \end{cases},
\end{equation}
where we assume $\eta_0 = 10^{14}\,\mathrm{cm}^{-2}\,\mathrm{s}^{-1}$ and $j_{\mathrm{cr}} = 250\,\mathrm{Fr}\,\mathrm{cm}^{-3}\,\mathrm{s}^{-1}$. The current threshold has been chosen so as to avoid Ohmic heating before the onset of the instability and to permit the ideal build-up to the instability. With this assumption, the minimum heating rate above threshold is $H = \eta_0 (4 \pi |j_{\mathrm{cr}}|/c)^2 \approx 0.3\,\mathrm{erg}\,\mathrm{cm}^{-3}\,\mathrm{s}^{-1}$, which is nearly 3 order of magnitude larger than the background heating ($H_0$) that we use for supporting the initial equilibrium. 
Below the critical current, a minimum numerical resistivity is inevitably present, but it does not provide any perceptible heating during the simulation.

\subsection{The loop setup}

Our computational 3D box contains two flux tubes of length $5 \times 10^9\,\mathrm{cm}$ and initial temperature of approximately $10^6\,\mathrm{K}$. Their footpoints are anchored to two thick, isothermal chromospheric layers at the top and bottom of the box. As the plasma $\beta$ decreases farther from the boundaries, the magnetic field expands. The initial atmosphere results from a preliminary simulation in which a box with a vertical magnetic field is relaxed to equilibrium until the maximum velocity falls below $10\,\mathrm{km}\,\mathrm{s}^{-1}$, as described in \cite{2014A&A...564A..48G}.

The computational box is a 3D Cartesian grid, $-x_M < x < x_M$, $-y_M < y < y_M$, and $-z_M < z < z_M$, where $x_M = 2 y_M = 8 \times 10^8\,\mathrm{cm}$, $y_M = 4 \times 10^8\,\mathrm{cm}$, and $z_M = 3.1 \times 10^9\,\mathrm{cm}$, with a staggered grid.
In the corona, we consider a non-uniform grid whose resolution degrades with height. To resolve the transition region at sufficiently high resolution, the cell size there ($|z| \approx 2.4 \times 10^9\,\mathrm{cm}$) decreases to $\Delta r \sim \Delta z \sim 3 \times 10^6\,\mathrm{cm}$ and remains constant across the chromosphere. Boundary conditions are periodic at $x = \pm x_M$ and $y = \pm y_M$, and reflective but with reversed sign for the tangential magnetic field component (equatorial symmetric boundary conditions) at $z = \pm z_M$.
\begin{figure}
\centering

\includegraphics[scale=0.25]{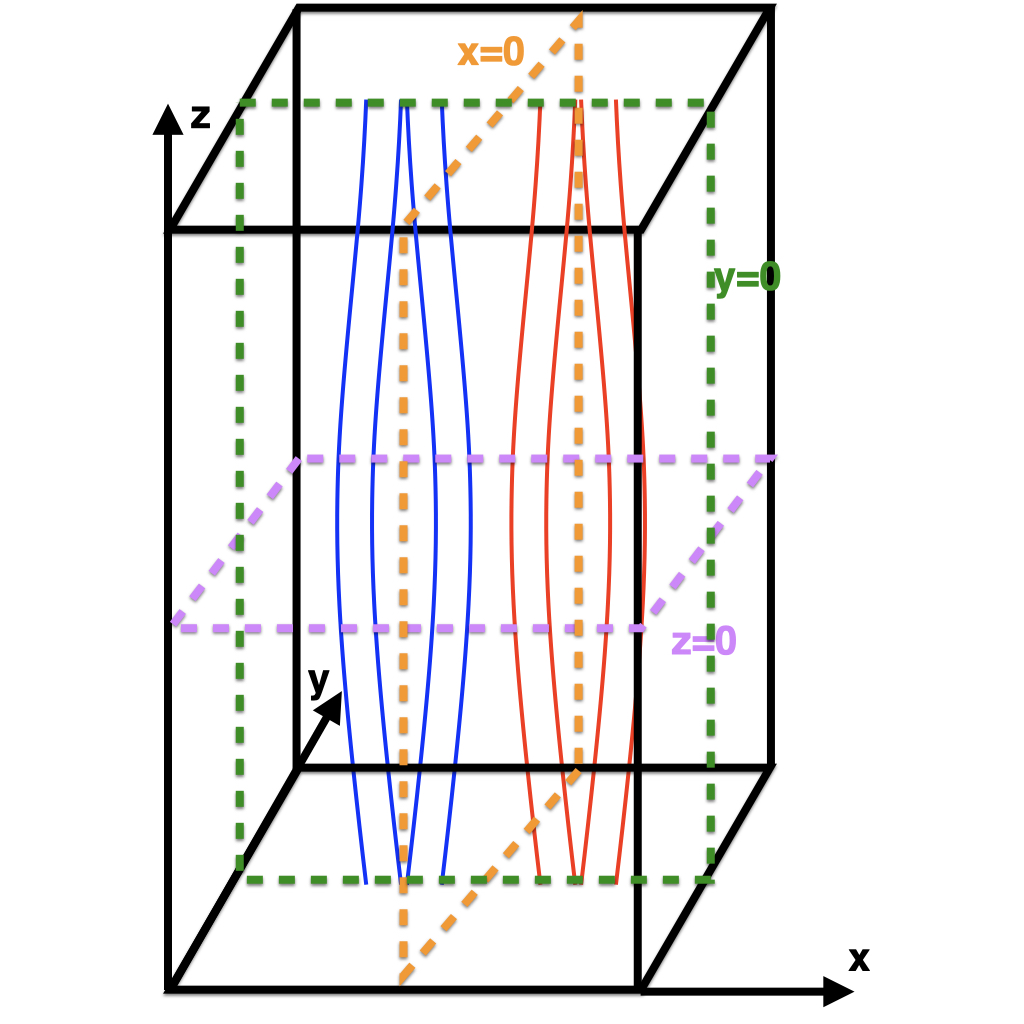}

\caption{Sketch illustrating the domain of the 3D simulation where to set of field lines (blue and red) are used to describe the initial positions of the magnetic flux tubes. The sketch is not to scale and it also shows some key planes for the analysis of the results.}
\label{fig:domainsketch}
\end{figure}
 
\subsection{Loop driver}

To trigger magnetic reconnection at the centre of the coronal domain, we use the upper and lower boundary conditions of the MHD model to devise a velocity transient driver for the chromospheric footpoints that slowly drifts them in opposite directions.
The aim is to generate a slight x-type misalignment in the corona, by imposing a footpoint y-velocity at the chromosphere defined as 
\begin{equation}
\label{equation_driver}
v_y(t)=\pm V_{max}\sin(2\pi\frac{t}{T_{driver}})
\end{equation}
for $t\le T_{driver}/2$, where $V_{max}=20~Km/s$ and $T_{driver}$ are chosen so the driver produces a $4^{\circ}$ footpoint displacement relative to the domain centre at $t=T_{driver}/2\sim360~s$.
The tilted geometry of the two flux tubes follows from the choice of signs in Eq.\ref{equation_driver}: at the lower boundary the $+$ sign applies for $x>0$ and $-$ for $x<0$, while at the upper boundary the signs are inverted.

\section{Results} \label{sec:results}

In this section we summarise the main results of the numerical investigation of the properties of nanojets and nanoflares. 
We first revisit a reference simulation, already presented in \citet{2021NatAs...5...54A}, which provides new insight on the mechanisms leading to magnetic reconnection, plasma heating, outward propagating jets, and evaporation into the coronal domain.
We then discuss other simulations where we change key parameters to study how the heating (nanoflare) and outward jets (nanojets) properties change. 
Namely, we consider different values for i) $j_{cr}$, the threshold current for magnetic resistivity, to analyse how nanoflares and nanojets change with different levels of magnetic stress, and
ii) the size of the simulation domain, to study how the energy released and jet speed change in smaller magnetic strands.

\subsection{Reference simulation}
\label{sec:ref250}

\begin{figure}
\centering

\includegraphics[scale=0.20]{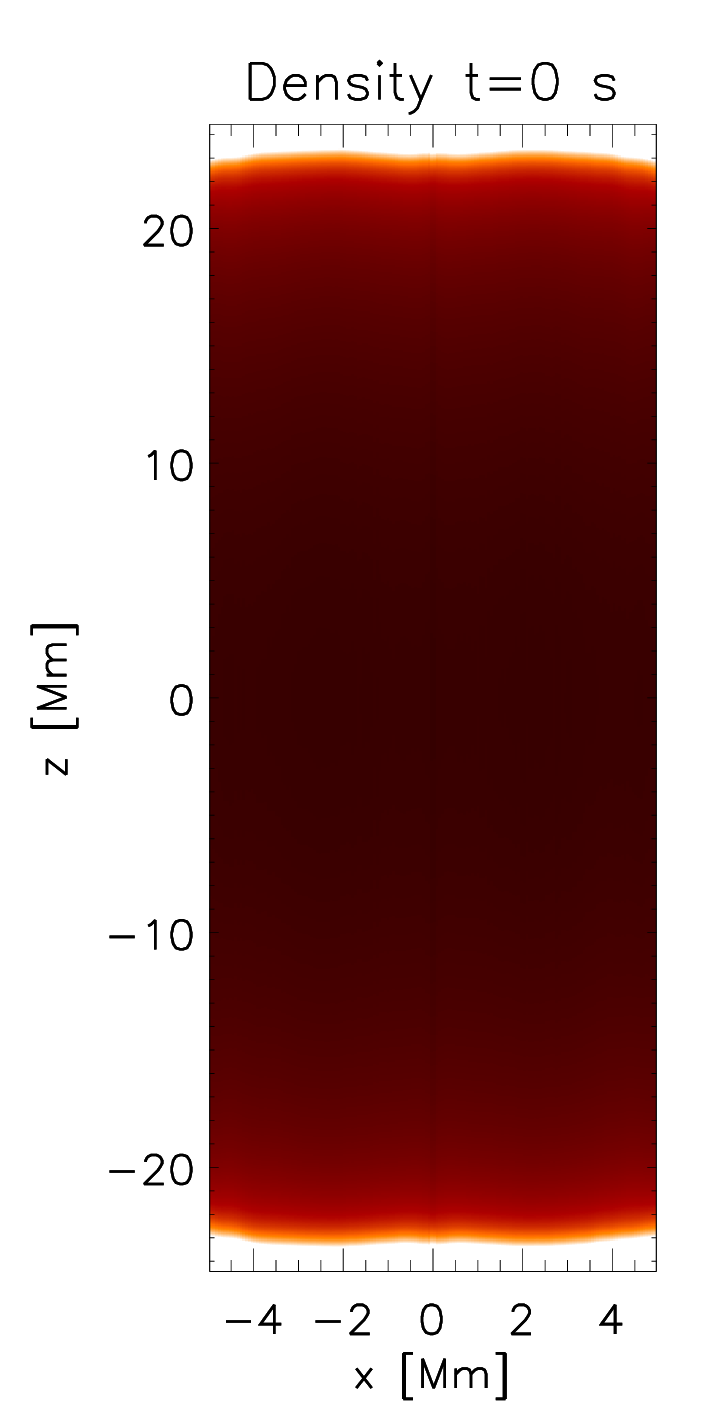}
\includegraphics[scale=0.20]{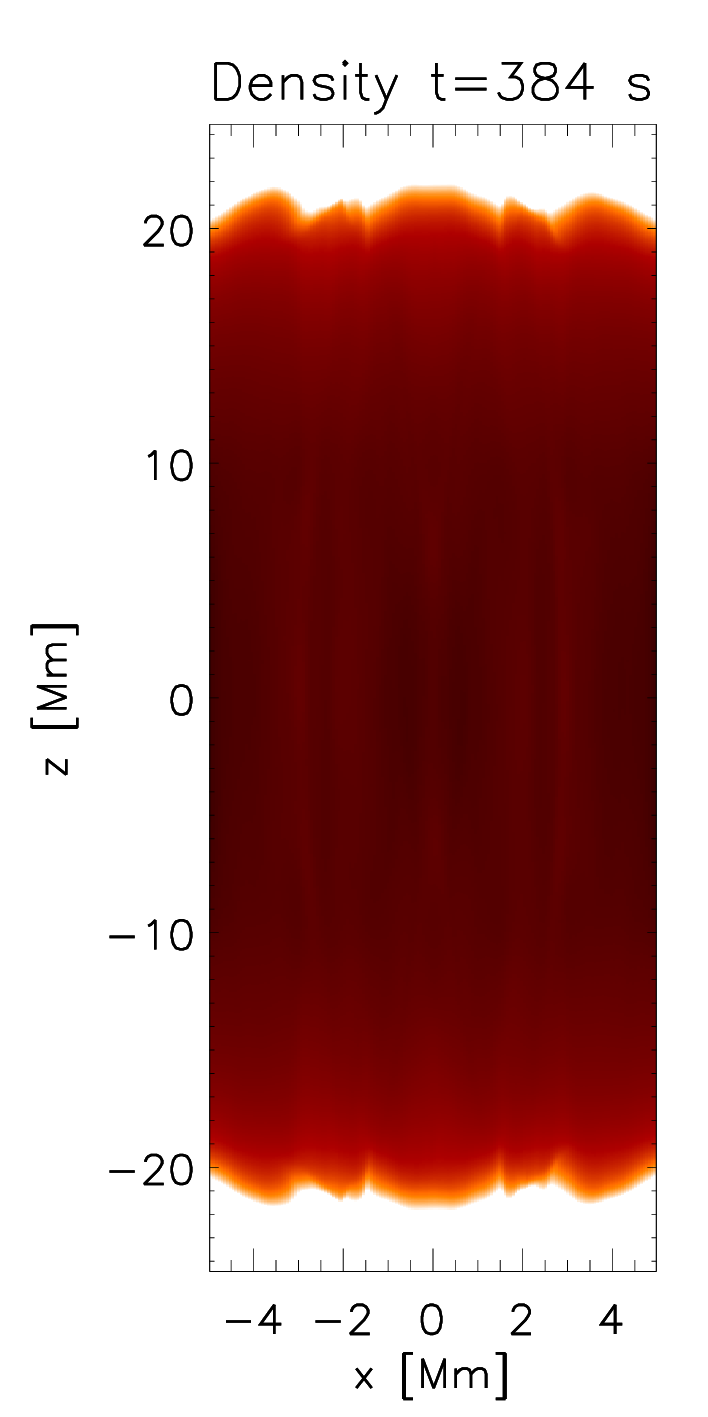}
\includegraphics[scale=0.20]{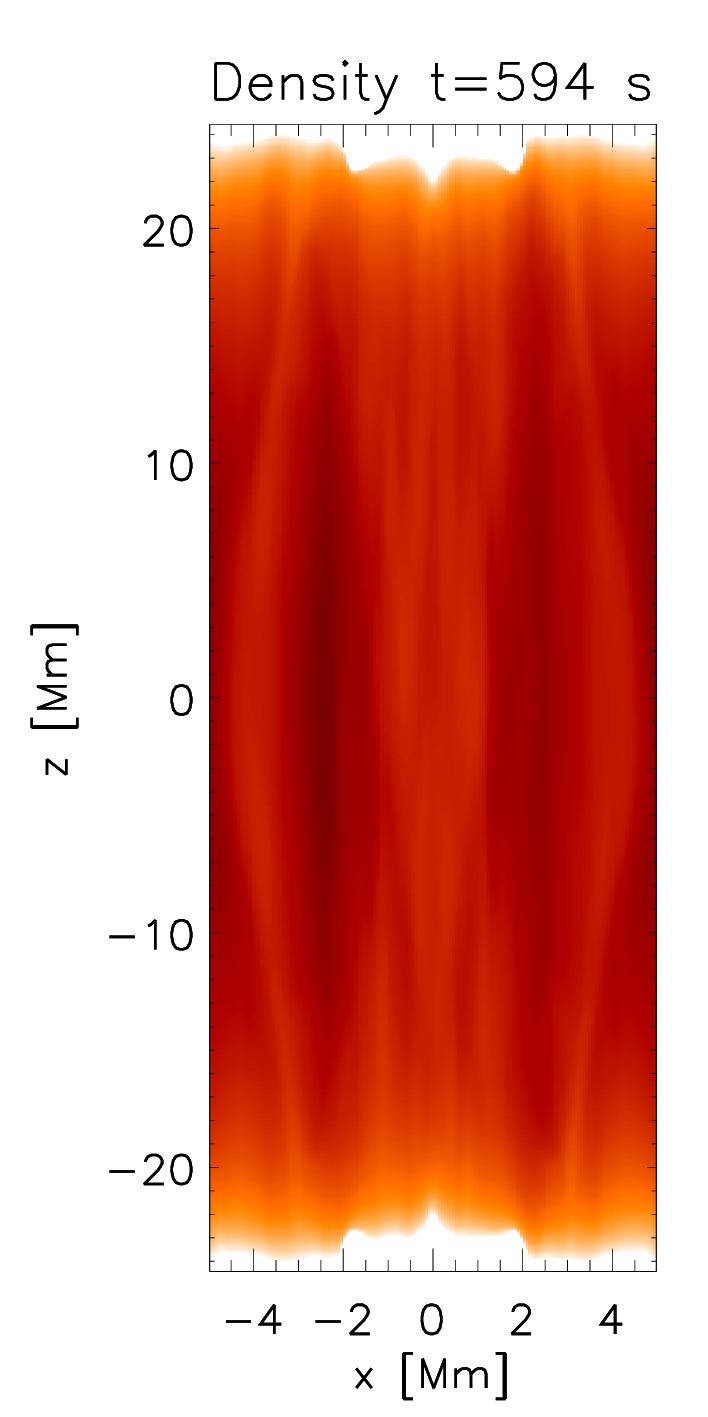}
\includegraphics[scale=0.20]{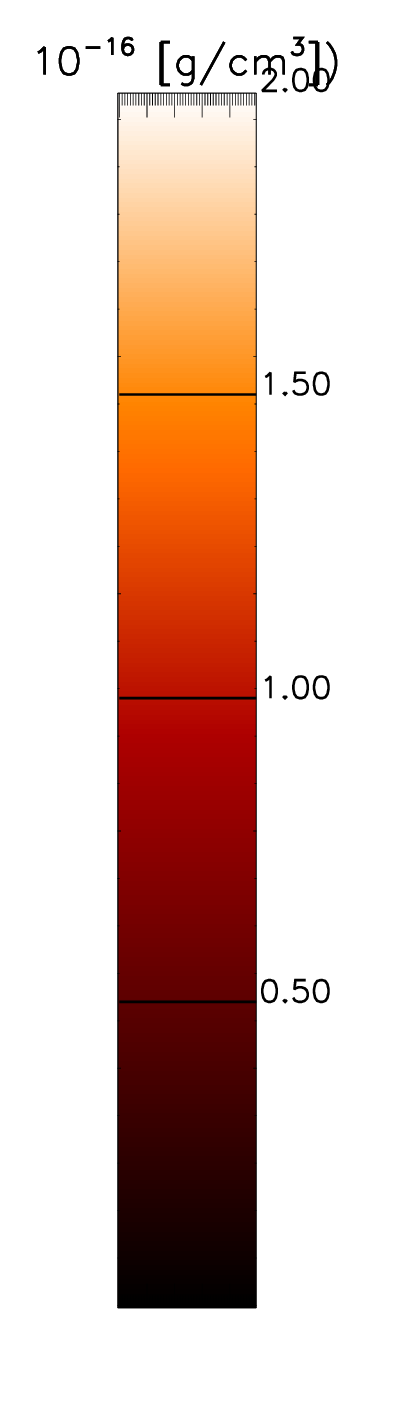}

\includegraphics[scale=0.20]{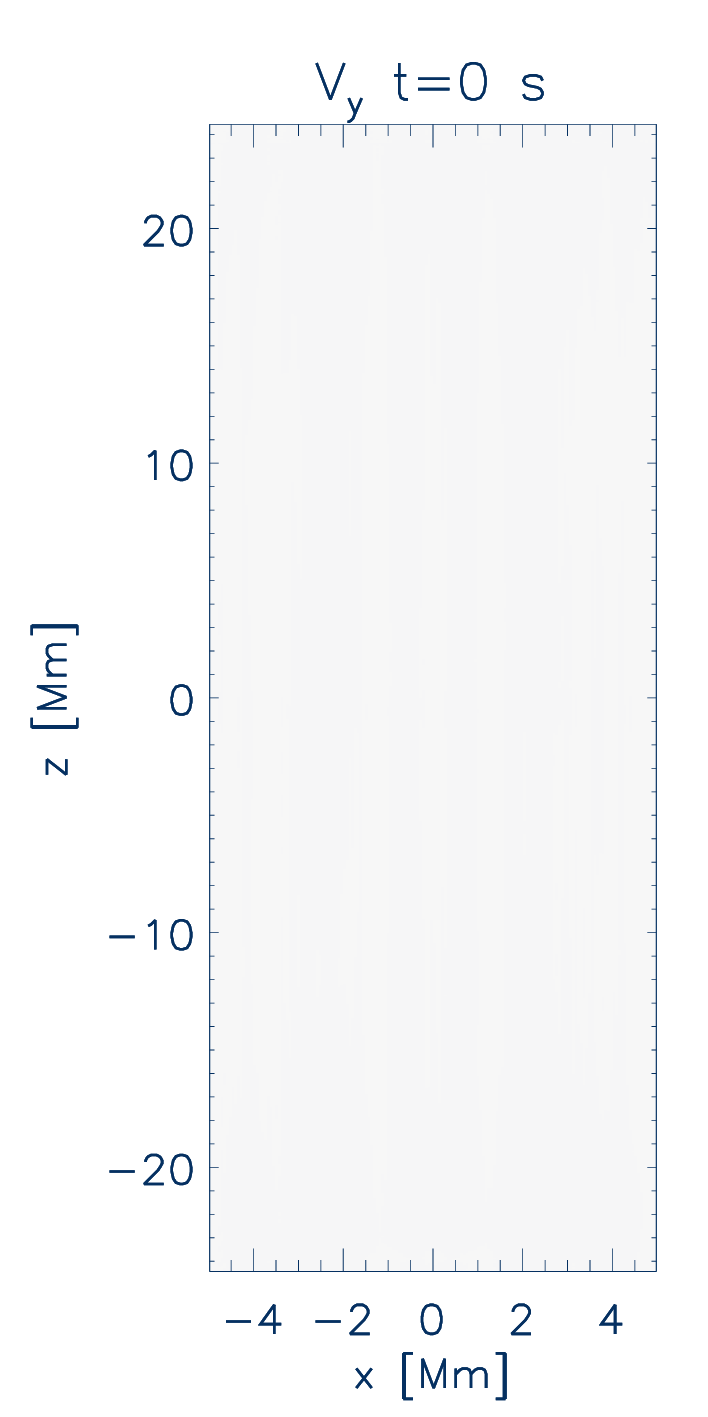}
\includegraphics[scale=0.20]{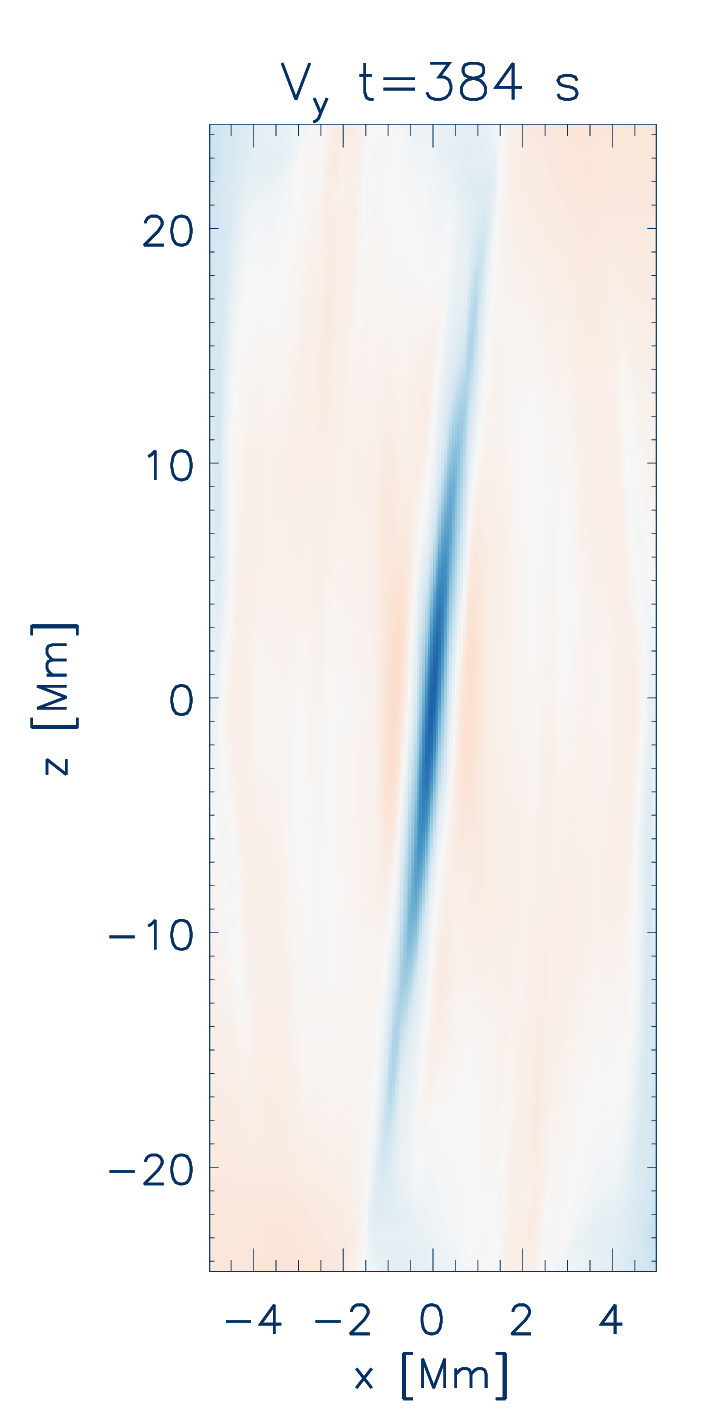}
\includegraphics[scale=0.20]{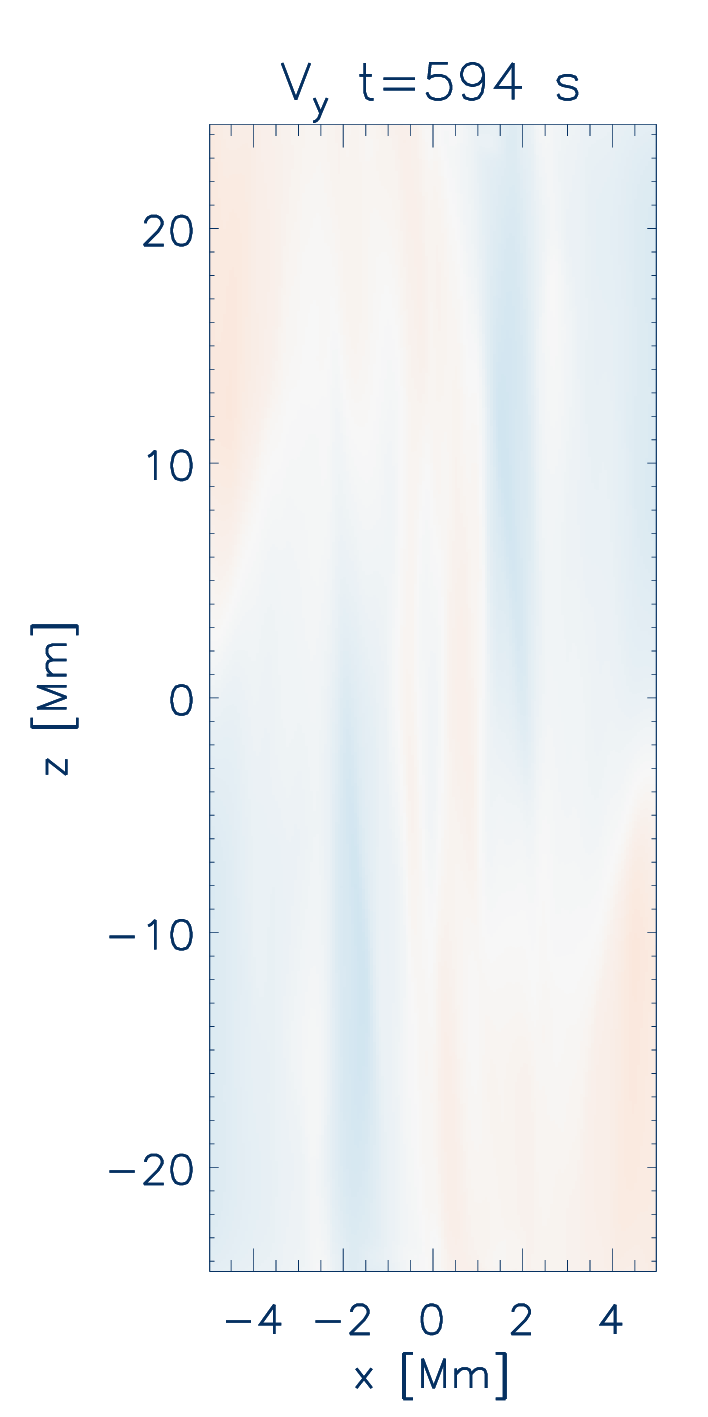}
\includegraphics[scale=0.20]{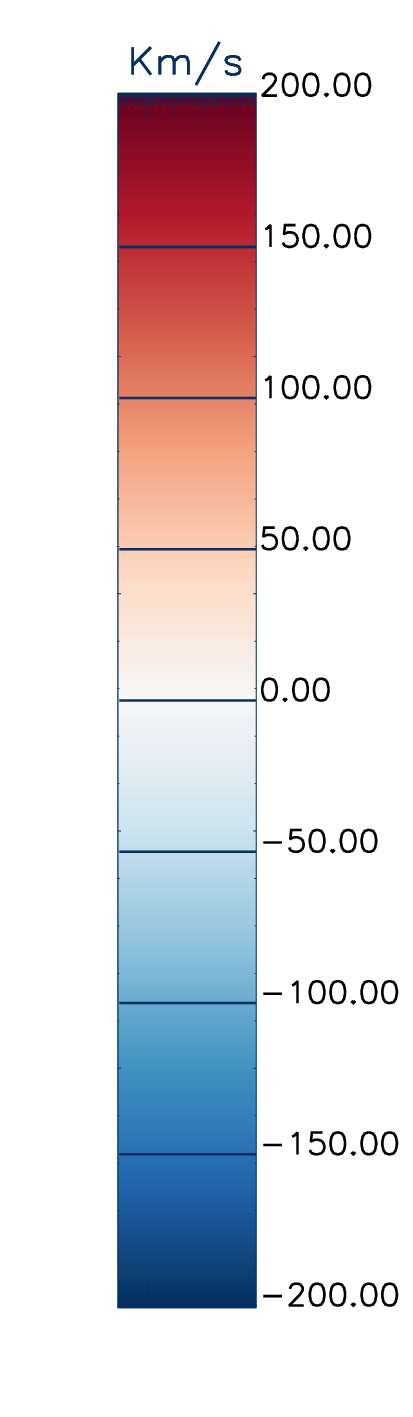}

\includegraphics[scale=0.20]{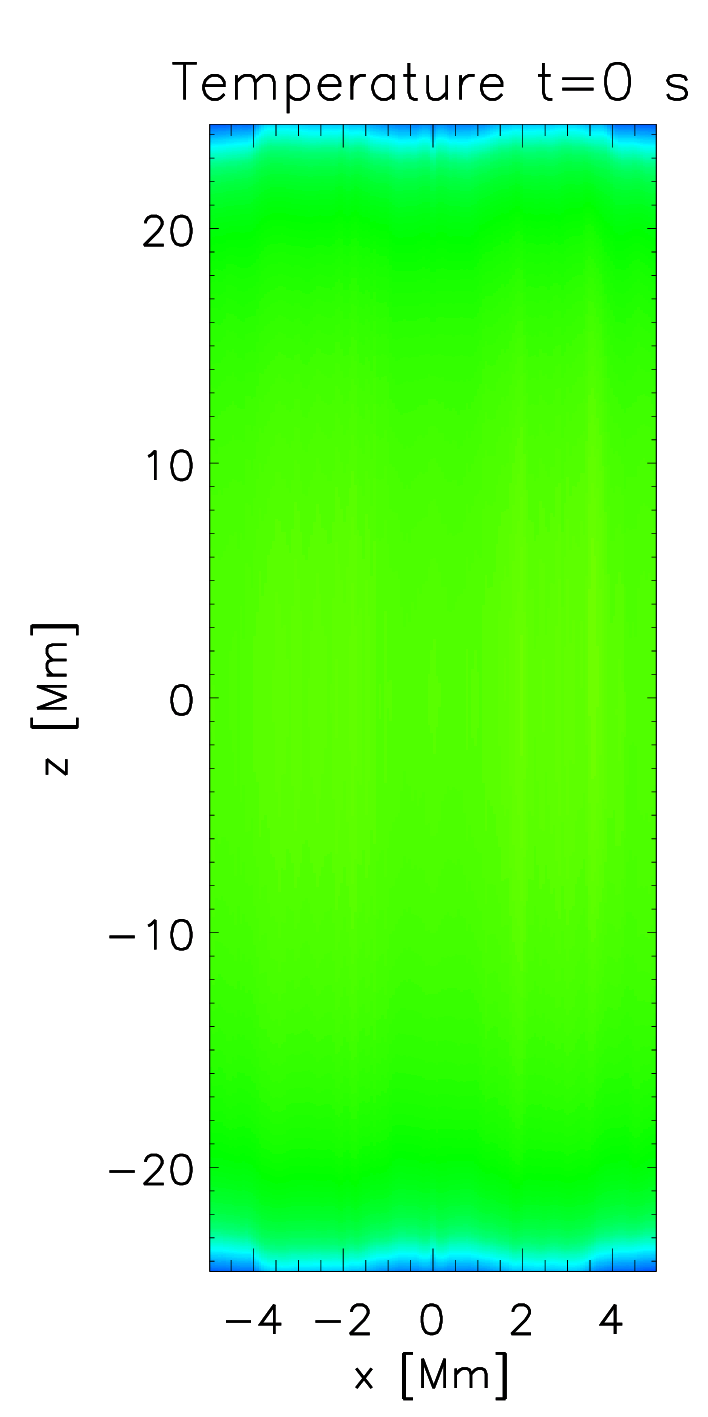}
\includegraphics[scale=0.20]{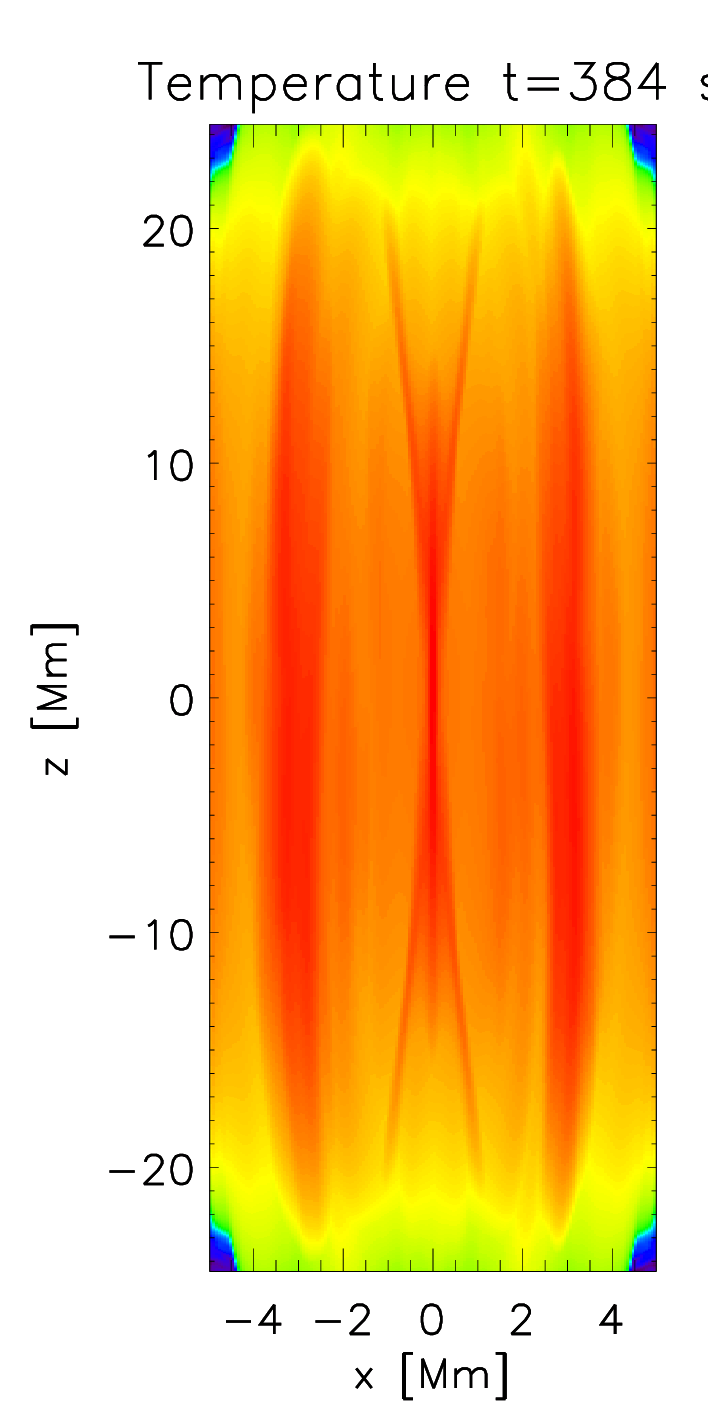}
\includegraphics[scale=0.20]{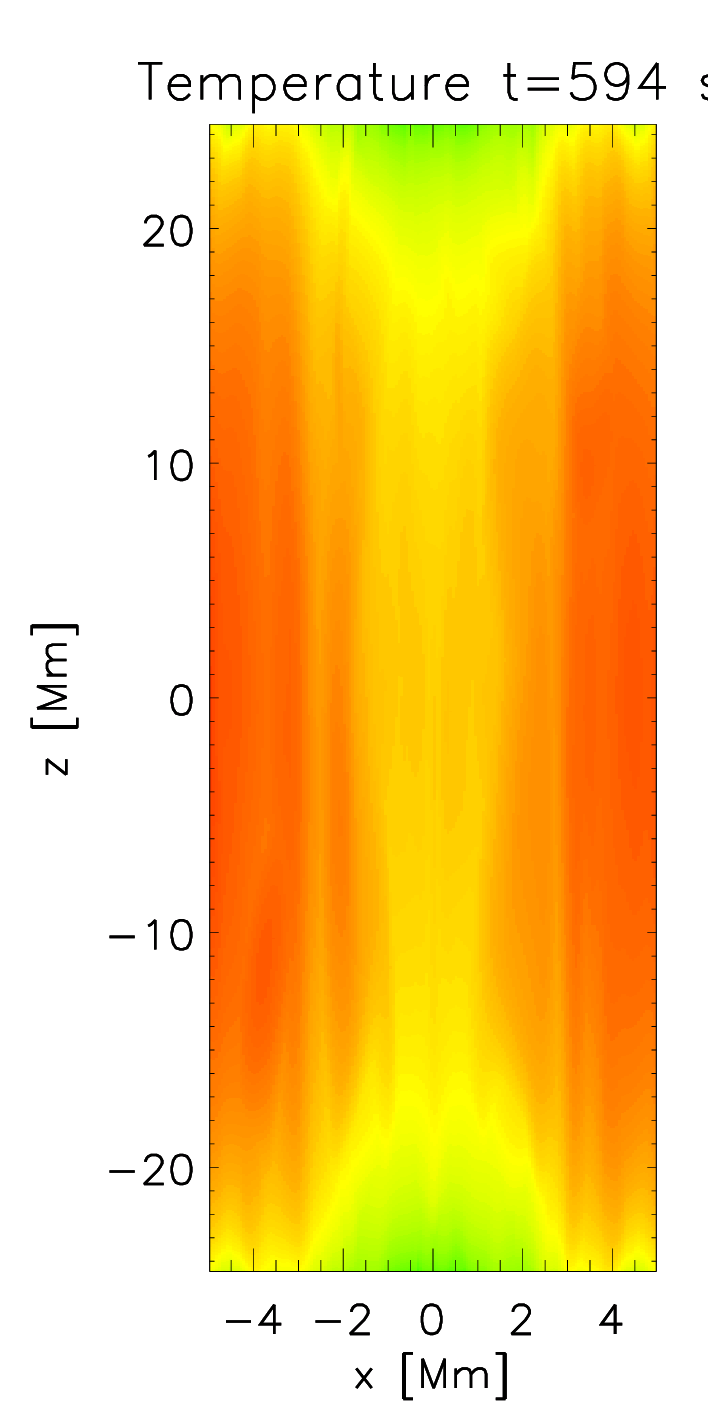}
\includegraphics[scale=0.20]{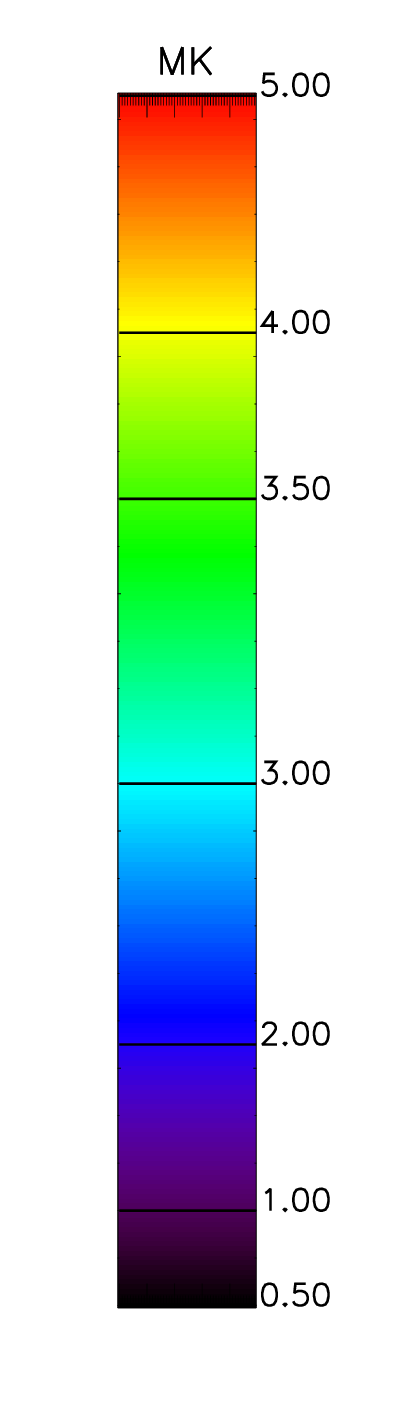}
\caption{Rendering of the average density, y-component of the velocity, and maximum temperature at $t=0$ (initial condition), $t=384~s$ (jets time), and $t=594~s$ (end of the simulation)}
\label{fig:rendering_j250}
\end{figure}

Figure \ref{fig:rendering_j250} shows the 3D rendering of density, y-component of velocity, and temperature onto the $x-z$ plane at three times, viewed along $y$ (aligned with the initial footpoint motion and the reconnection jets).
The density maps show the average density along $y$; the velocity maps show the y-component of velocity averaged by density for $y\le0$, avoiding overlap with opposite velocities at $y>0$; the temperature maps show the maximum temperature along each line of sight.
Once the driver displaces the footpoints along $y$, the coronal magnetic fields of either flux tube become increasingly tilted at the separation surface $x=0$.
This triggers reconnection at the centre of the domain and two jets form; our rendering shows only the one for $y\le0$ (Fig.\ref{fig:rendering_j250}, $t=384-s$).
At this time the density only increases slightly at the 
footpoints and it corresponds to mild compression from the displacement of the initial structures, with a mass density increase below $25\%$.
The temperature instead increases significantly. Part of this heating (at $x=\pm3~Mm$) is caused by such compression: footpoint motion compresses the neighbouring plasma, pushing it upwards from either footpoint, and when these flows meet at the apex plane ($z=0$) they compress and heat further.
At the same time, the temperature reaches up to 5.5 MK at the domain centre, with similarly high temperatures at the flux-tube centres ($x=\pm3$ Mm).
This ohmic heating is only active very close to the $x=0$ plane, so the temperature increase elsewhere in the domain must be ascribed to plasma compression (the hot flanks near $x\pm2~Mm$ in the temperature map of Fig.\ref{fig:rendering_j250} at $t=384~s$).
At $t=384~s$ the jets are already formed near the domain centre (Fig.\ref{fig:rendering_j250}), with a y-velocity exceeding $300~km/s$; these velocity components extend in $z$ along the curvature of the newly reconnected field lines.
Once reconnection is exhausted and both jet and heating have stopped, the density is significantly higher, with coronal mass increased by a factor of 3.
The temperature has dropped to about 2 MK by then and keeps decreasing, especially at the centre, once heating stops.

\subsection{Magnetic reconnection trigger}

An important aspect we address here is what magnetic configuration builds up the conditions for reconnection.
In our view a key feature of braiding is to drive the system towards reconnection. Here, an initial equilibrium is evolved into a configuration where reconnection is the way to minimise the free magnetic energy. This happens upon the effect of a footpoint motion that makes the system slip to reconnection.

\begin{figure}
\centering

\includegraphics[scale=0.25]{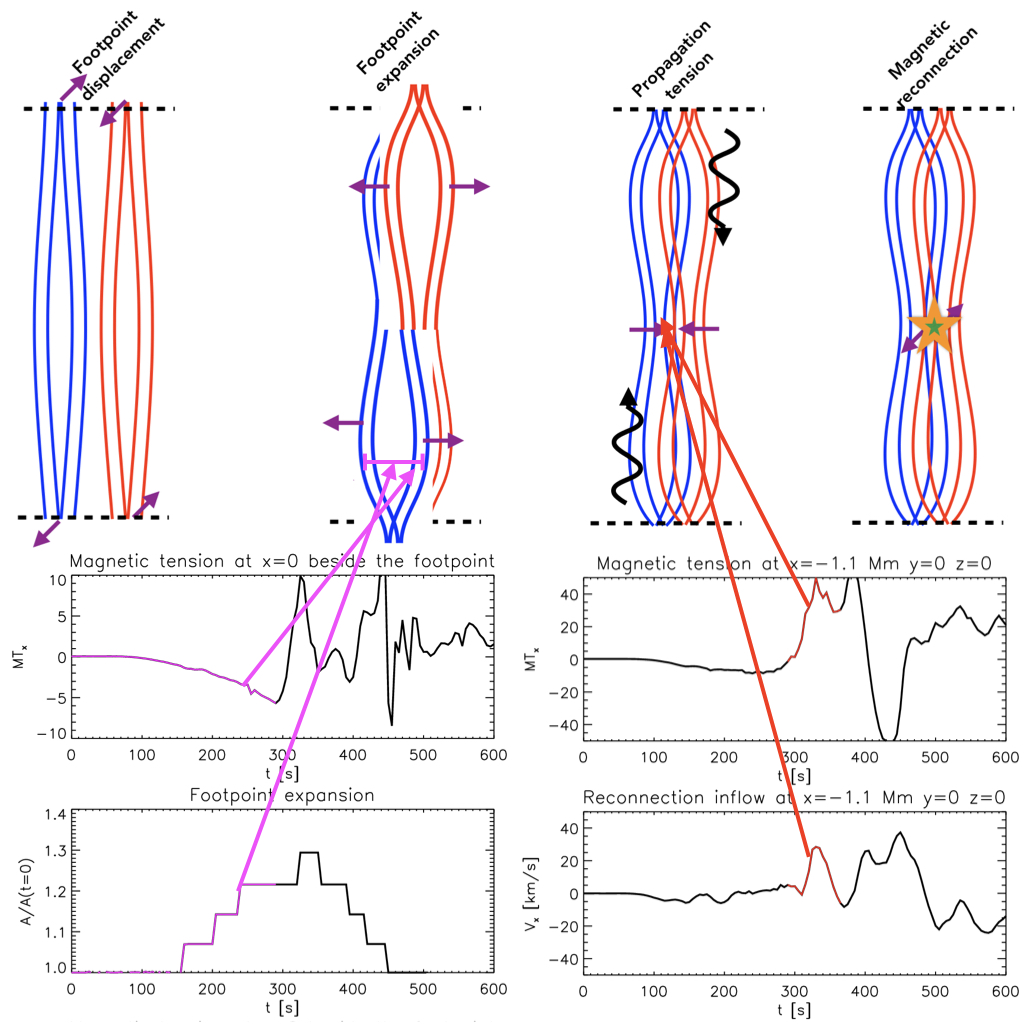}
\caption{Top panels: sketch illustrating the flux tubes dynamic leading to the magnetic reconnection. It represents four different phases of the relative position and configuration of the two flux tubes, identified by a sketch of their magnetic field lines (blue and red).
Mid panels: the left panel shows the evolution of the magnetic tension at $x=0$, at the (moving) y-coordinate of the lower footpoint of the left-hand-side flux tube; the right panel shows the evolution of the magnetic tension near the centre of the domain, at the point $(x,y,z)=(-1.10,0,0)~Mm$.
Bottom panels: the left panel shows the expansion of the cross-section area of the flux tube that encloses the initial magnetic flux; the right panel shows the plasma velocity near the centre of the domain, at the point $(x,y,z)=(-1.10,0,0)~Mm$.
In these panels, the segments of the curve highlighted in magenta and red mark the time interval of the evolution that is relevant for the sketch in the top panels as also indicated by the arrows.}
\label{fig:triggerreconnection}
\end{figure}

The sketch in Fig.\ref{fig:triggerreconnection} illustrates the underlying mechanism in our simulation.
The initial configuration, with the two flux tubes parallel, is distorted by the boundary driver displacing the footpoints in opposite directions, one moving away from the other.
When the driver stops at $t=175~s$, the footpoints are farthest apart, and the flux tubes expand near the footpoints as the magnetic pressure excess is no longer balanced by the neighbouring tube.
This is an inevitable consequence of perturbing the initial equilibrium where the flux tubes slip one beside the other. Were the perturbation much smaller or in the x-direction, the flux tubes could have found a new equilibrium without triggering the mechanism leading to magnetic reconnection.

This is shown in the second panel of Fig.\ref{fig:triggerreconnection} where, after an initially negligible expansion, the area of the cross section of the flux tube that encompasses the initial magnetic flux grows to about 1.2 by the end of this braiding phase.
This is corroborated by the insurgence of magnetic tension near the footpoints, directed inward towards the centre of the expanded flux tubes, as shown quantitatively by the evolution of the magnetic tension at the position $x=0$ moving in the y-direction with the footpoint of the left flux tube at the lower boundary: it increases in intensity during this initial phase, pointing in the negative direction, as suggested by the sketch.

The expansion at either footpoints then bends the field lines so that the magnetic tension also pushes for an expansion at the centre of the flux tube, pushing the field lines of the two tubes toward each other.
This is shown in the evolution of the magnetic tension at $x=-1.12~Mm$ and $z=0$, near the reconnection location, which becomes positive, pushing towards the current sheet as the footpoint motion advances.
This magnetic tension insurgence travels from the footpoints towards the centre covering a distance of about $20~Mm$ in a few tens of seconds for propagation speed of about $1000~km/s$, close to the local Alfv\'en speed of about $1300~km/s$ from a magnetic field strength of $\sim 15~G$ and a density of $\sim 10^{-15}~g/cm^3$ , consistent with a sausage mode propagating along the flux tube in a low-$\beta$ regime.
When the magnetic tension enhancement reaches the centre and pushes the plasma and frozen-in field lines from either tube to converge, this is the ultimate trigger for reconnection.
The plasma flows towards the reconnection point at about $20~km/s$, as shown by the evolution of $v_x$ at $x=-1.12~Mm$, $z=0$ and 
such a convergent motion builds an electric current at a rate of $4.5~G/s^2$.
Once the current threshold for magnetic resistivity is met, reconnection is set and the plasma is slingshotted away from the domain centre at $v\sim300~km/s$.
In summary, reconnection is triggered by magnetic forces from the expansion of the flux tubes, propagating from the footpoints after they move away from an unstable equilibrium.
The idea of an unstable equilibrium at the onset of reconnection was already explored in other works \citep[e.g.][]{2010A&A...516A...5W, 2011A&A...525A..57P, 2015ApJ...808..134C}

\subsection{Nanojet and nanoflare}

\begin{figure*}
\centering

\includegraphics[scale=0.19]{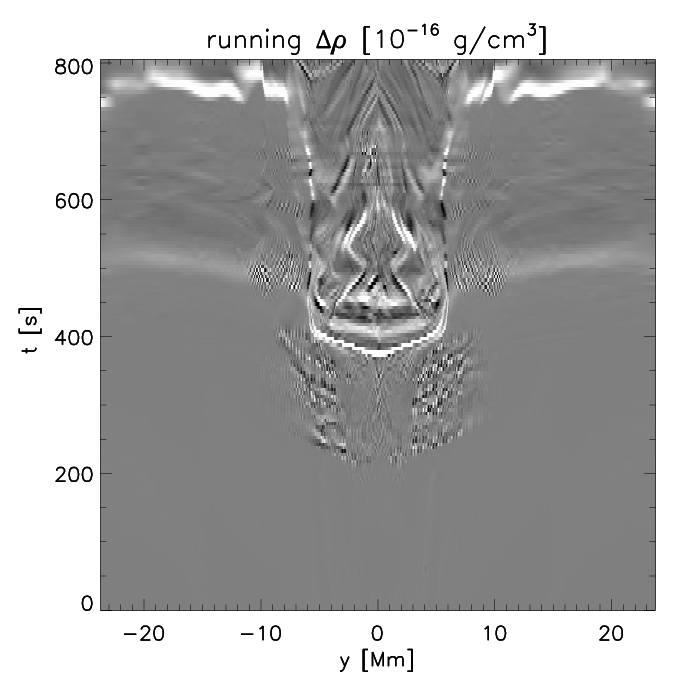}
\includegraphics[scale=0.19]{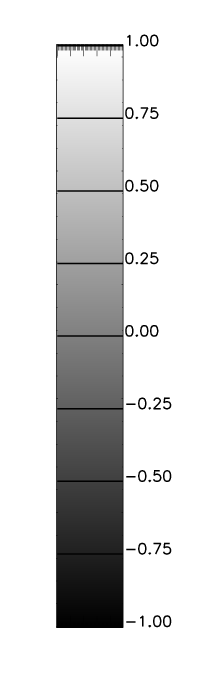}
\includegraphics[scale=0.19]{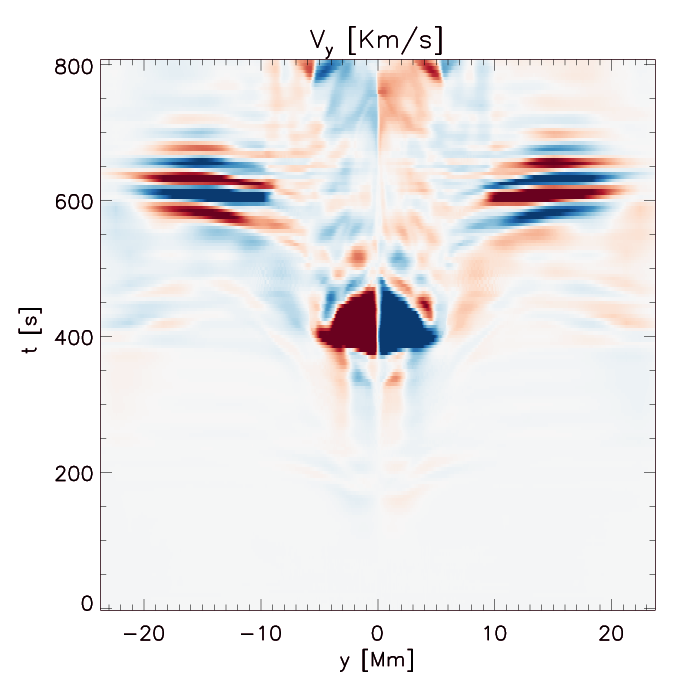}
\includegraphics[scale=0.19]{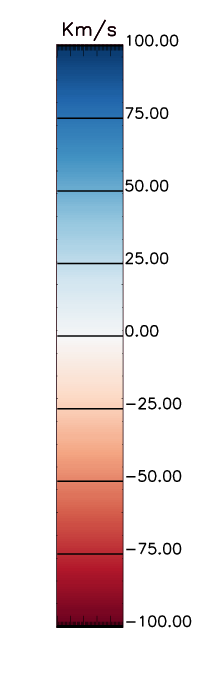}
\includegraphics[scale=0.19]{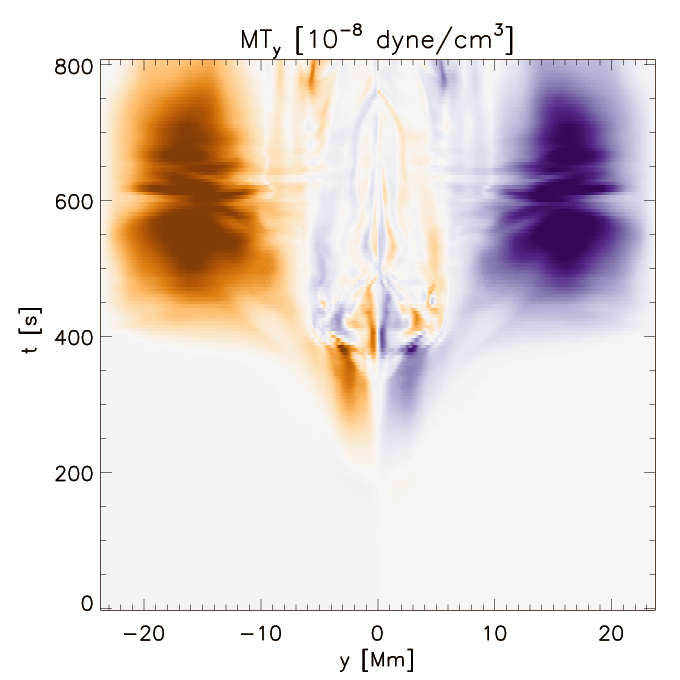}
\includegraphics[scale=0.19]{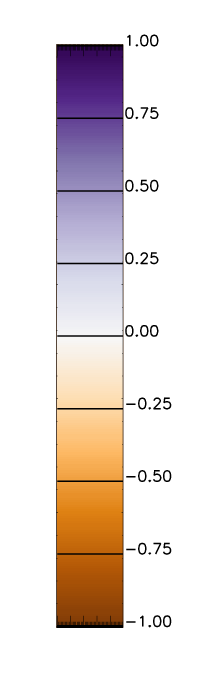}

\includegraphics[scale=0.19]{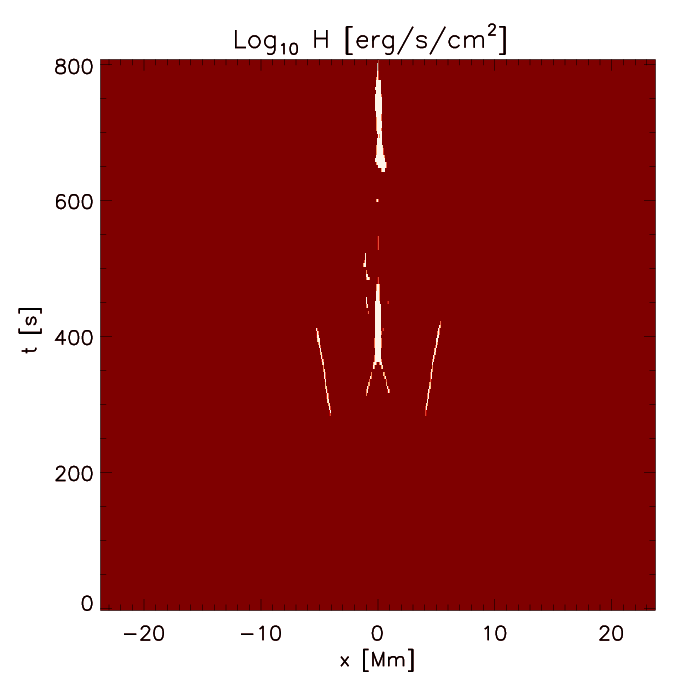}
\includegraphics[scale=0.19]{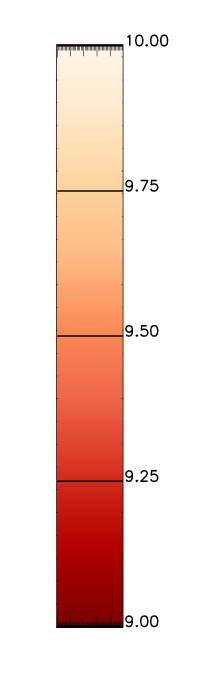}
\includegraphics[scale=0.19]{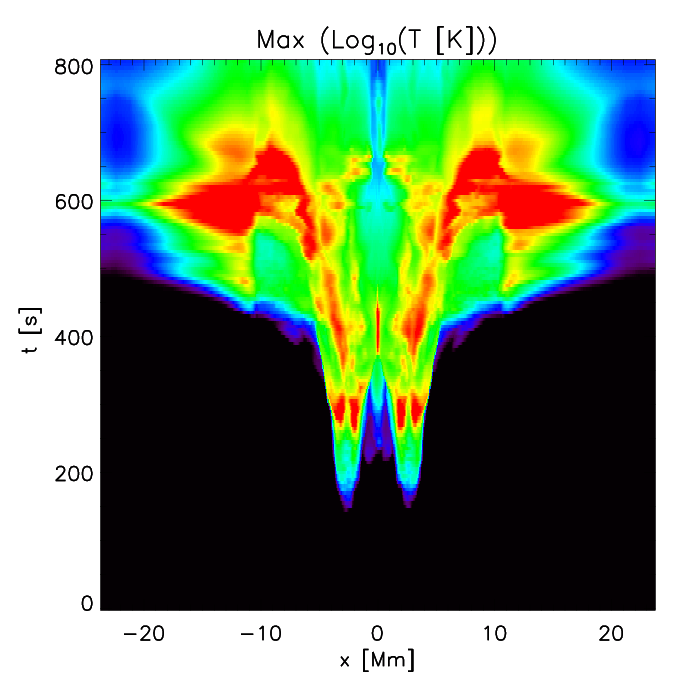}
\includegraphics[scale=0.19]{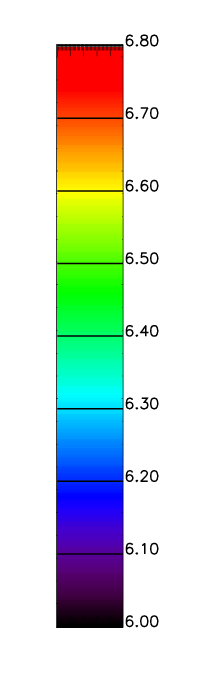}

\caption{Time distance maps of various quantities along the central cut of the simulation going across the initial position of the flux tubes ($z=0$, $y=0$). The shown quantities are density running difference,  $v_y$, the y-component of the magnetic tension, the logarithm of the ohmic heating, and the logarithm of the maximum of the temperature along the y-direction.}
\label{fig:tdmaps_j250}
\end{figure*}

It is interesting to explore the relationship between reconnection heating and its dynamic counterpart, the jet.
To illustrate this, we consider time-distance maps along the $x=0$, $z=0$ cut. Fig.\ref{fig:tdmaps_j250} shows these maps for running differences of density, y-velocity, y-component of magnetic tension, heating, and temperature.
Reconnection kicks in at $t=365~s$, and from then on the density running differences clearly show plasma suddenly moving outwards in both directions, caused by the magnetic tension pushing the plasma (see the magnetic tension panel in Fig.\ref{fig:tdmaps_j250}). Following the density front, we derive an outward motion of about $180~km/s$, while plasma speed in the jet region reaches $\sim300~km/s$. The density front then suddenly stops around $t=420~s$, having travelled about $7~Mm$.
This sudden stop occurs when plasma motion has curved the field lines enough to generate a restoring magnetic tension opposing the jet motion, which develops rather rapidly, after $\sim30~s$, at about $4~Mm$ from the reconnection point.
These patches of inverted magnetic tension correspond to y-velocity patches opposite to the jets. The plasma-$\beta$ stays of order $10^{-2}$ throughout, so this restoring mechanism is mostly magnetic rather than due to plasma compression.
Later, oscillatory structures appear in the y-velocity time-distance maps and this mechanism naturally triggers transverse waves.

From a heating perspective, ohmic heating is released at the domain centre over a short period of about $100~s$.
In our model, ohmic heating can occur only where the current density exceeds the threshold ($j_c=250~G/s$), after which heat propagates along the field lines.

\begin{figure}
\centering

\includegraphics[scale=0.25]{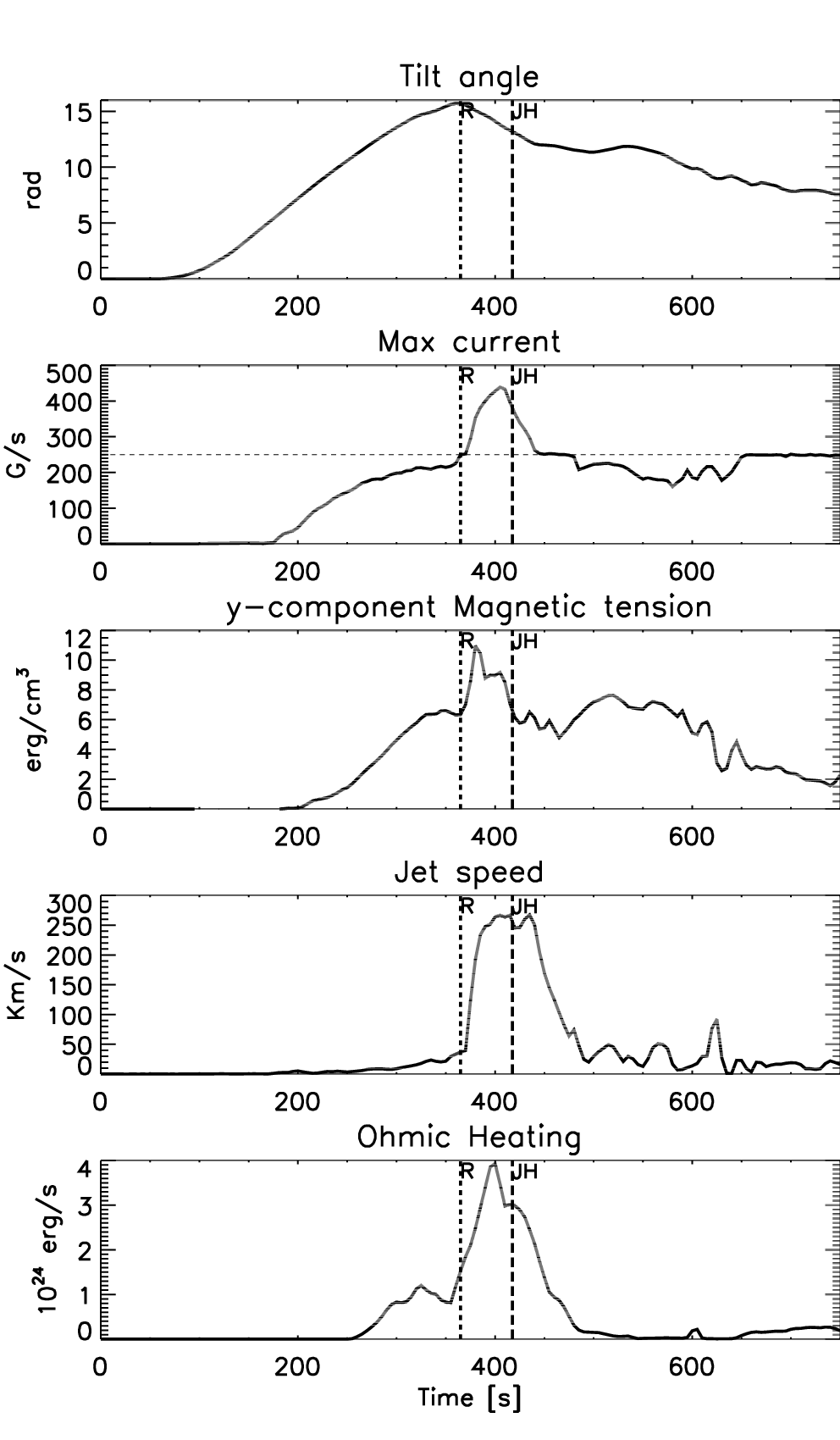}

\caption{Evolution as a function of time of the tilt angle between the two magnetic flux tube, the maximum electric current between the two flux tubes, the y-component of the magnetic tension at the jet locations, the maximum jet speeds, the heating at the reconnection location, and the mass evaporation from under the corona.}
\label{fig:evolhis}
\end{figure}
In Fig.\ref{fig:evolhis} we plot key quantities illustrating the development of reconnection and its consequences.
From top to bottom, the tilt angle between the axial fields of the two loops steadily increases between $t=0$ and $t=380~s$, when it starts to decrease.
The electric current between the two loops, piles up until it reaches the threshold that triggers magnetic resistivity at $t=365~s$ — the time we identify as the start of reconnection.
The second panel follows the maximum electric current in the central cut ($x=0$, $z=0$), which ramps up to even larger values, while the maximum magnetic tension along the same cut (third panel) rapidly increases. The maximum $|v_y|$ along the same cut (fourth panel), representing jet speed, similarly ramps up because of the magnetic tension.
At the same time, the ohmic heating (the integral of $\eta j^2$ between $z=-20~Mm$ and $z=20~Mm$) rises and peaks during reconnection.

The magnetic tension suddenly increases as field lines reconnect due to magnetic resistivity, triggering jets in two opposite directions. 
At the same time, the jets advect magnetic flux and pressure away from the reconnection site, producing an inward total pressure gradient that slightly accelerates reconnection further.

At $t=405~s$ the thrusting effect of the jets ends and the electric current starts decreasing. Reconnection is still ongoing, so heating and high velocity persist untile $t=475~s$.
The loop plasma temperature changes continuously throughout, with the largest increase during reconnection.

\subsection{Accumulation of magnetic stress}

\begin{figure}
\centering
\resizebox{\hsize}{!}{\includegraphics[scale=0.32]{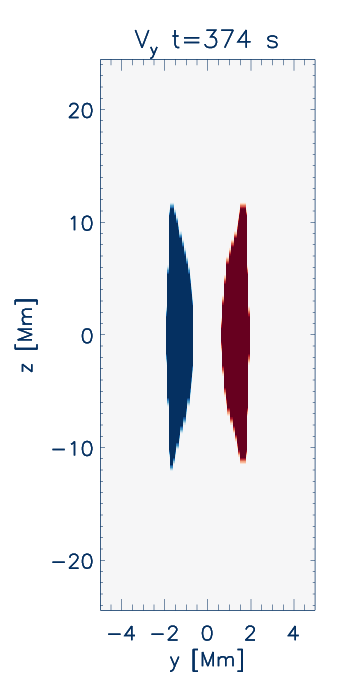}
			        \includegraphics[scale=0.32]{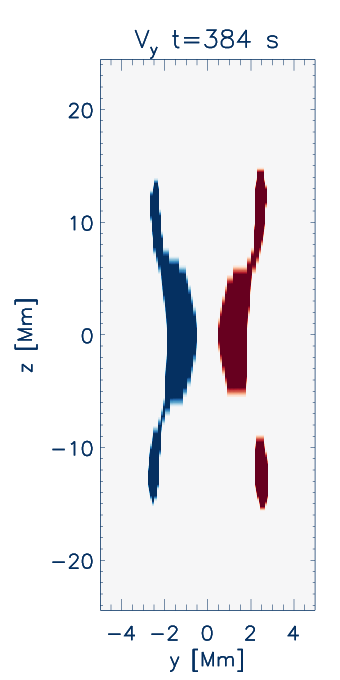}
			        \includegraphics[scale=0.32]{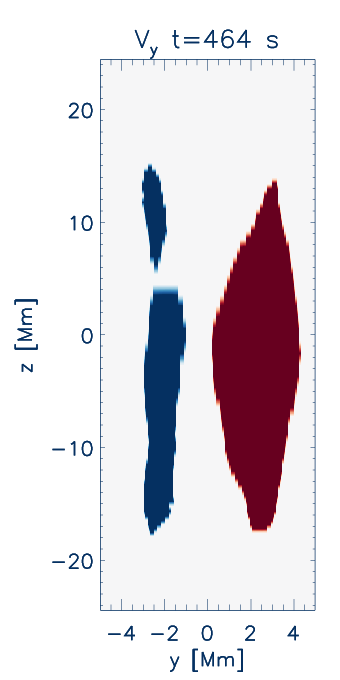}
			        \includegraphics[scale=0.32]{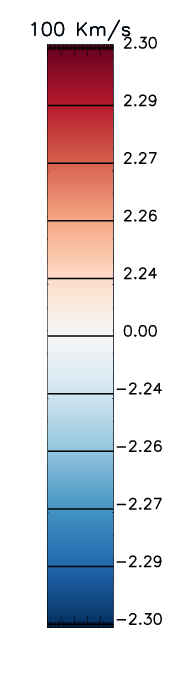}}
\caption{Rendering 3D of Vy in the y-z plane for different simulations.
$j_{\mathrm{cr}} = 100\,\mathrm{Fr}\,\mathrm{cm}^{-3}\,\mathrm{s}^{-1}$, $j_{\mathrm{cr}} = 250\,\mathrm{Fr}\,\mathrm{cm}^{-3}\,\mathrm{s}^{-1}$, $j_{\mathrm{cr}} = 500\,\mathrm{Fr}\,\mathrm{cm}^{-3}\,\mathrm{s}^{-1}$.}
\label{fig:vy_xz_jcrit}
\end{figure}

We investigate how jet properties and energy release change with different amounts of accumulated magnetic stress — key for the nanoflare scenario, or specifically for coronal-loop braiding, where braiding must consolidate before an energy burst.
With higher $j_c$, the current threshold triggering magnetic resistivity, the two loops become increasingly tilted relative to each other.
This tilt leads to increasing electric currents between the two loops, dissipated only once the threshold current is met. The threshold value thus determines when the tilt stops growing, currents start dissipating, temperature starts rising, and the field starts relaxing.
We compare simulations with a low current threshold ($j_{\mathrm{cr}} = 100\,\mathrm{Fr}\,\mathrm{cm}^{-3}\,\mathrm{s}^{-1}$, left panel of Fig.\ref{fig:vy_xz_jcrit}), our reference simulation ($j_{\mathrm{cr}} = 250\,\mathrm{Fr}\,\mathrm{cm}^{-3}\,\mathrm{s}^{-1}$, central panel), and a high threshold ($j_{\mathrm{cr}} = 500\,\mathrm{Fr}\,\mathrm{cm}^{-3}\,\mathrm{s}^{-1}$, right panel).
We omit units for $j_c$ in the rest of the text for simplicity.
In Fig.\ref{fig:vy_xz_jcrit}, for each simulation we display the time of maximum y-velocity around the reconnection point, to identify the jet's maximum speed.
The figure shows a rendering of the y-velocity weighted by plasma density along the line of sight, selecting only pixels with $v_y\ge 200~km/s$, projected on the $x-z$ plane to view the jets from the side.

As expected, jets are triggered later as the current threshold increases, which also leads to stronger magnetic tension and thus higher outward jet speeds.
The simulations with $j_c=100$ and $j_c=250$ mainly differ in the shape of the reconnection jets, while maximum velocity stays similar; at $j_c=250$ we measure only about 20\% more total jet momentum, but the two are overall very similar. 
In contrast, $j_c=500$ shows much larger jets, significantly higher speed, and a developing asymmetry.

\begin{figure}
\centering

\includegraphics[scale=0.25]{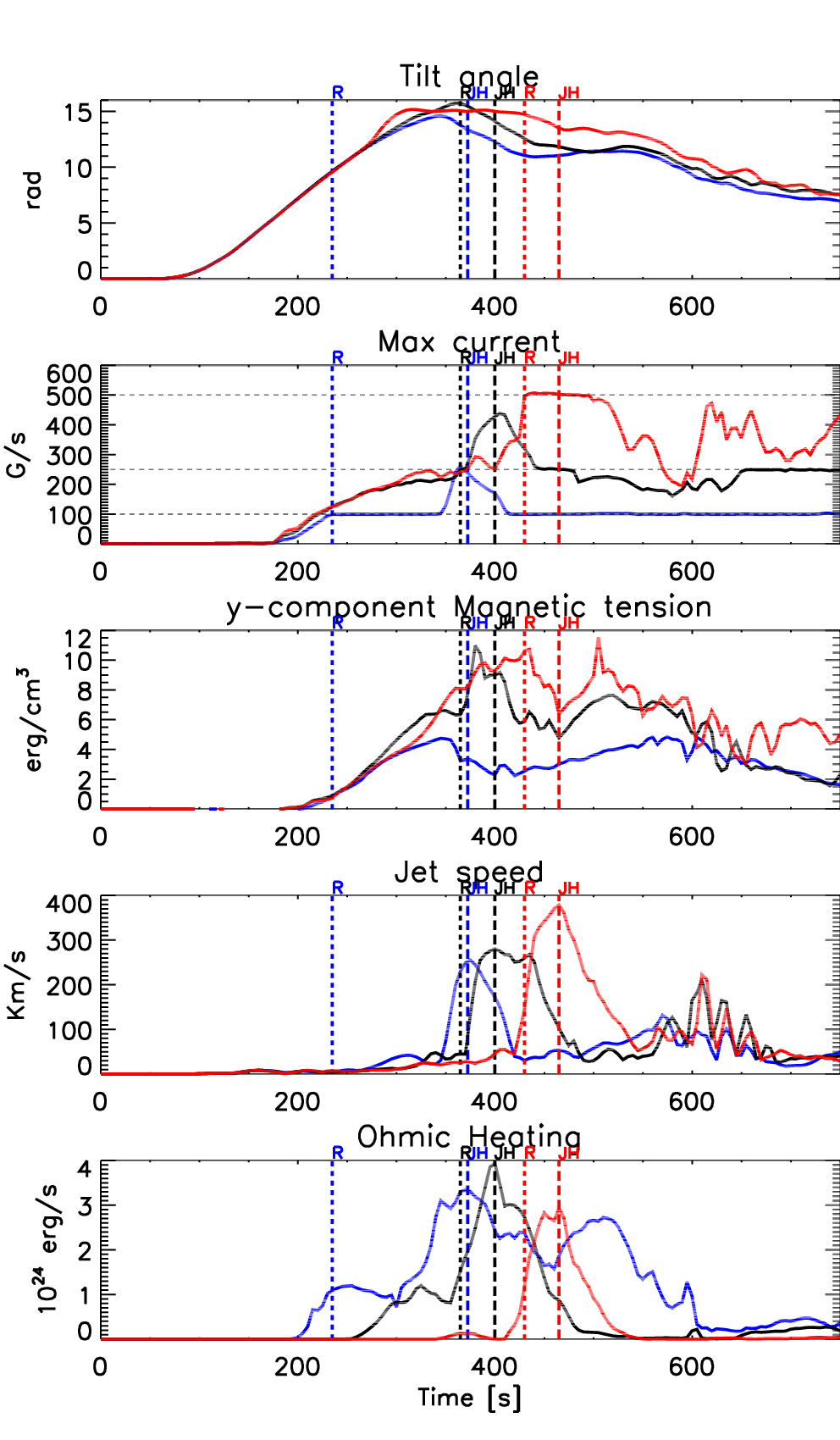}

\caption{Evolution of the same quantities as in Fig.\ref{fig:evolhis} for the three simulations with $j_c=100$ (blue lines), $j_c=250$ (black lines), and $j_c=500$ (red lines).}
\label{fig:simhissome}
\end{figure}

In Fig.\ref{fig:simhissome} we show the evolution of the same quantities as in Fig.\ref{fig:evolhis} (Sec.\ref{sec:ref250}) for these three simulations, finding that different triggering conditions ($j_c$ values) lead to qualitatively different system evolution.
The tilt angle initially increases for all simulations; for $j_c=100$ and $j_c=250$ it simply stops growing and immediately unwinds. For $j_c=500$, instead, the tilt angle plateaus for about $150~s$ before decreasing.

For $j_c=100$ the critical current is reached at $t=235~s$ (dashed line R); afterward, the currents dissipate exactly at the critical value while the tilt angle is forced to increase. 
Later ($t=368~s$) the current jumps, triggering faster evolution, and the tilt angle starts decreasing.
For $j_c=250$, the maximum current jumps immediately after reaching threshold, and the tilt angle decreases as well.
For $j_c=500$, there is no current jump; it plateaus for about $100~s$, after the tilt angle has already stopped growing, before both current and tilt angle decrease together. 

Consequently, the magnetic tension responsible for the slingshot generating the jets evolves differently across the three models.
With the low threshold, reconnection starts early and the field immediately begins unwinding; the tension soon stops increasing and stays low. 
With higher thresholds, tension develops further, exerting a stronger force on the plasma in the reconnection region.
The tension peak is reached before the current peak, especially for $j_c=100$ and $j_c=250$, where the current bump follows the plasma displacement caused by the tension peak — not the case for $j_c=500$.

In terms of jet velocity, $j_c=100$ and $j_c=250$ show similar maximum speed, with $j_c=250$ slightly higher and more prolonged, while $j_c=500$ shows significantly higher maximum velocity ($\approx400~km/s$).
In this regime, a higher current threshold seems to lead to more magnetic stress accumulated during twisting, converting automatically into stronger jets.

The asymmetry that is well visible in the simulation with $j_c=500$ can be traced to a corresponding asymmetry in the y-component of the magnetic tension on either side of the reconnection region. This asymmetry is present in all the simulations, growing from about 10\% for $j_c = 100$ to 40\% for $j_c = 250$ and reaching 100\% for $j_c = 500$. This also explains why a localised velocity of $400~km/s$ appears in only one of the two outflows of the simulation. Also, the visible asymmetry in Fig.\ref{fig:vy_xz_jcrit} is partially a rendering effect, as the velocity levels chosen accentuate the contrast.

Both $j_c=100$ and $j_c=250$ show current peaks well above threshold, occurring after the plasma velocity has begun increasing due to magnetic tension, explained by shear currents generated by the fast plasma displacement carrying the field with it. 
There is thus a positive feedback: reconnection triggered by slow footpoint motion generates the jet, which produces stronger currents and more dissipation. 
We do not find this for $j_c=500$, where the threshold is too high for plasma-motion currents to be significant.

Things are different, however, when we focus on the heating. 
Fig.\ref{fig:simhissome} shows the total integrated coronal heating, revealing two competing effects when changing the current threshold.
A lower threshold lets a larger coronal volume exceed it, enabling ohmic heating, though the heating stays close to $H_c=\eta j_c^2$.
Higher $j_c$ thus means smaller heated volumes but higher local heating values.
The cumulative heating slowly rises for $j_c=100$, peaks, then decreases without vanishing until $t=600$ s. For $j_c=250$, heating persists between $t=250$ s and $t=500$ s — starting later due to the higher current required, but stopping earlier as the current drops below threshold sooner; its peak, however, is higher than for $j_c=100$.
The $j_c=500$ simulation shows the shortest heating duration and the lowest peak.

\begin{figure}
\centering

\includegraphics[scale=0.38]{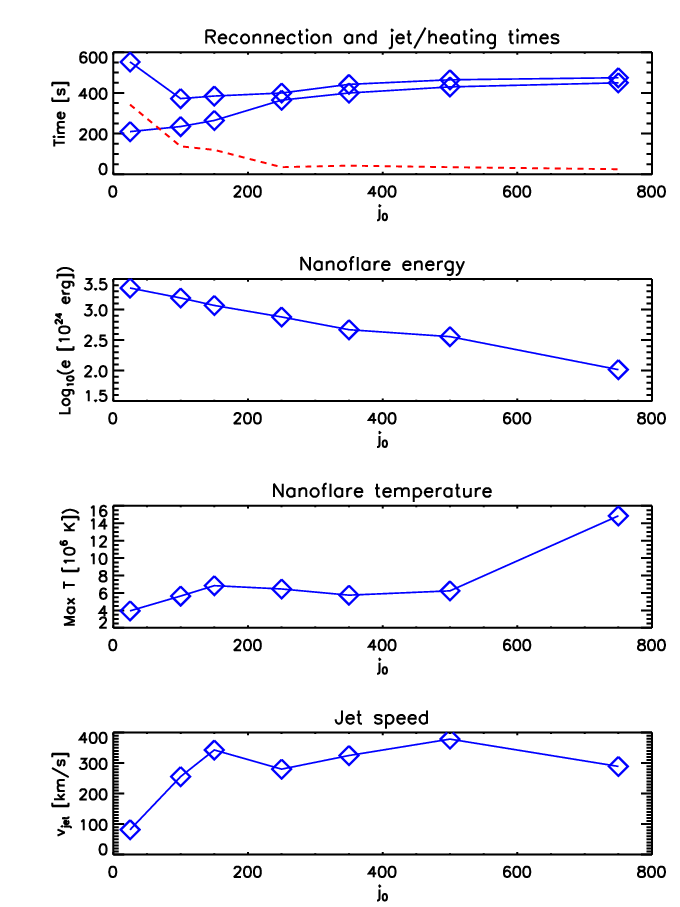}

\caption{Time of the occurrence of the nanojet, energy of the nanoflare, maximum temperature and maximum speed of the nanojet as a function of the simulation parameter $j_c$. The red dashed lines in the top panel displays in seconds the time difference between the reconnection time and the peak of the jet speed and heating time.}
\label{fig:jcrit_time_energy_speed}
\end{figure}

We then ran simulations varying $j_c$ over $25, 100, 150, 250, 350, 500, 750$; Fig.\ref{fig:jcrit_time_energy_speed} summarises the results.
The top panel shows the time the current threshold is reached as a function of $j_c$, and the time of the nanojet/nanoflare (the average between the times of maximum velocity and heating).
These times differ by about $300~s$ for small $j_c$, but the two events become nearly simultaneous as $j_c$ increases.
The lack of secondary currents generated by reconnection at high $j_c$ makes the two events coincide, though the transition between regimes is smooth.

The energy released by ohmic heating decreases monotonically with $j_c$: the lower $j_c$, the easier ohmic heating occurs and the more energy converts to thermal energy. 
The $j_c=750$ simulation has fast jets but little heating, as few points exceed the critical current. At the other extreme, $j_c=25$ allows continuous current dissipation and larger ohmic heating, but never enough magnetic tension buildup for a slingshot strong enough to produce a jet.
While heating from the reconnection event clearly decreases with $j_c$, the maximum nanoflare temperature shows more complex behaviour, increasing from $j_c=25$ to $j_c=150$ and then very slowly decreasing.
Jets become significant only for $j_c\ge100$, with speeds near $300~km/s$; this maximum speed is roughly the same across simulations regardless of $j_c$, and the behaviour is not even monotonic.
This can be explained by the interaction between the jet and the background magnetic field and plasma, an obvious limit on outward plasma travel independent of $j_c$.

\begin{figure}
\centering
\includegraphics[scale=0.38]{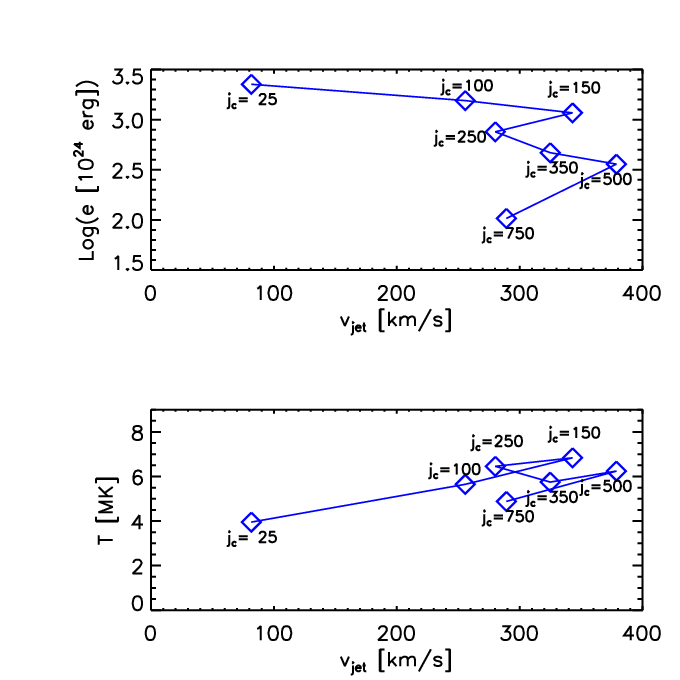}
\caption{Enery of the nanoflare (upper panel) and maximum temperature (lower panel), as a function of the speed of the nanojet.}
\label{fig:speed_energy_temperature}
\end{figure}

Combining this analysis, we can examine how ohmic heating and nanoflare temperature depend on nanojet velocity, as shown in Fig.\ref{fig:jcrit_time_energy_speed}.
These plots show no obvious monotonic relation between heating and velocity — between the nanoflare's heating and its dynamic counterpart.
It thus does not seem viable to infer a clear dependence between jet speed and heating amount, as shown in Fig.\ref{fig:speed_energy_temperature}.
Only some basic considerations can be drawn: simulations with higher $j_c$ generally group around higher velocity, slightly higher temperatures, and lower heating.
On the other hand, higher heating rates are associated with both slow nanojets ($j_c=25$, $90~km/s$) and faster ones, up to $j_c=150$ at $350~km/s$.
In terms of temperature, all nanojets between $250~km/s$ and $400~km/s$ group in mixed order around $4$–$8~MK$.

\subsection{Energy of the nanoflare}
Our simulations show total ohmic heating several orders of magnitude larger than the typical nanoflare estimate of about $10^{24}$ erg, or even less, as found in \citet{2013ApJ...770L...1T}.

One possibility is that the simulations over-produce heating because the modelled system is large compared to the scale at which nanoflares typically occur.
To test this, we ran simulations reducing the domain size in $x$ and $y$ while leaving the vertical $z$ direction unchanged — effectively describing thinner loop strands without changing the flux tube length. 

For this we rerun the initial 3D relaxation, using current thresholds $j_c=250$ and $j_c=500$, halving the domain extent in $x$ and $y$ and thus reducing the flux-tube cross-section by a factor of 4.

\begin{figure}
\centering

\includegraphics[scale=0.26]{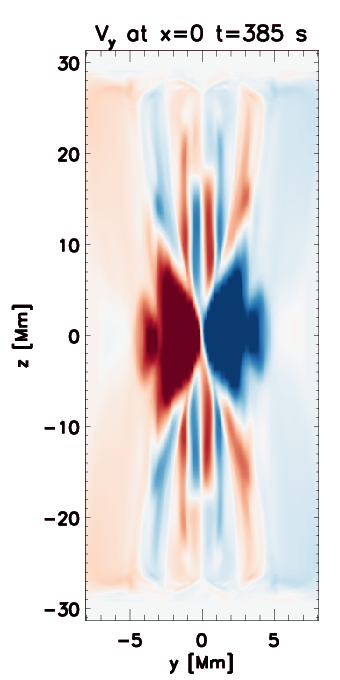}
\includegraphics[scale=0.26]{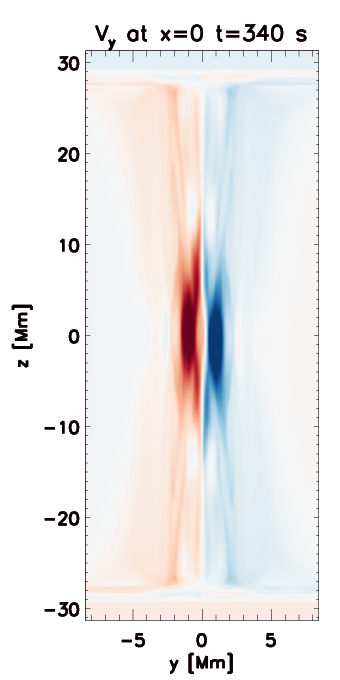}
\includegraphics[scale=0.26]{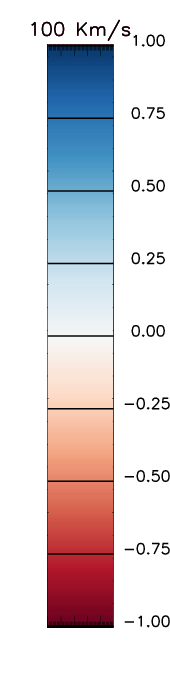}

\caption{Maps of the $y-$component of the plasma velocity on the y-z plane at $x=0$ at the times of the maximum jet speed for the reference simulation and the simulation in a smaller domain.}
\label{fig:small250vy}
\end{figure}

Fig.\ref{fig:small250vy} compares maps of the y-velocity at the $x=0$ plane for the two $j_c=250$ simulations at their respective jet times, with the reference simulation zoomed in to match scale.
Visually, the jets are very similar in shape, vertical extension, and speed, with both cases around $\sim100~km/s$.

However, the ohmic heating produced differs substantially.
\begin{figure}
\centering
\includegraphics[scale=0.32]{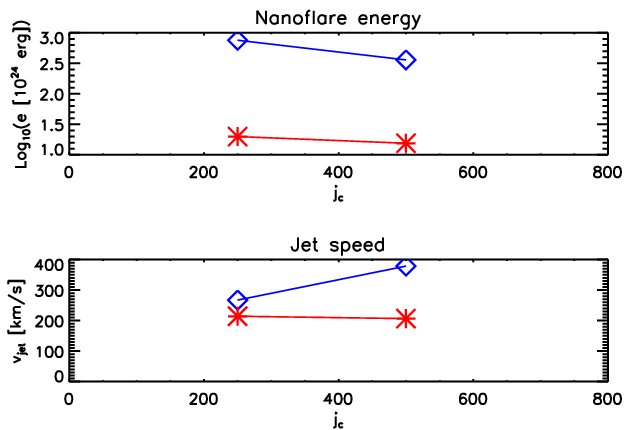}
\caption{Energy of the nanoflare and maximum speed of the nanojet as a function of the simulation parameter $j_c$ for simulations with either domain sizes. Blue squares represent the simulations with the reference domain and red stars represent the simulations with a small domain.}
\label{fig:smalljetvar}
\end{figure}
Fig.\ref{fig:smalljetvar} shows the heating and nanojet velocity for $j_c=250$ and $j_c=500$, reference (blue) versus small domain (red).
With smaller flux tubes, the maximum jet speed stays similar for $j_c=250$, but is significantly smaller (though same order of magnitude) for $j_c=500$.

The heating, however, differs greatly: it drops by about 2 orders of magnitude, to about $10^{25}~erg$, for both $j_c$ values.
This drop shows that our reconnection scenario can match the key nanoflare heating constraints — maintaining jet speeds of hundreds of $km/s$ while releasing energy much closer to $10^{24}~erg$.
This numerical experiment suggests reconnection inside coronal loops, at strand level, can indeed explain the nanoflare scenario.

It remains to be understood whether such small events can be detected, even deploying reconnection outflows at $\sim100~km/s$. Reconnection outflows have proved difficult to observe, so far possible only for a limited set of events and only with new-generation telescopes.

\subsection{Emission synthesis} \label{sec:lines}

We are also interested in the detectability of nanojet and nanoflare features, key to understanding how much of this analysis future solar-corona instruments could address.

We focus on EUV emission in two iron lines: Fe\,XV ($\lambda=284.16~\AA$, $T_{\rm max}=2.5$~MK) and Fe\,XIX ($\lambda=108.36~\AA$, $T_{\rm max}=10$~MK), both intense and selected for the forthcoming MUSE mission.

\begin{figure}
\centering

\resizebox{\hsize}{!}{\includegraphics[scale=0.45]{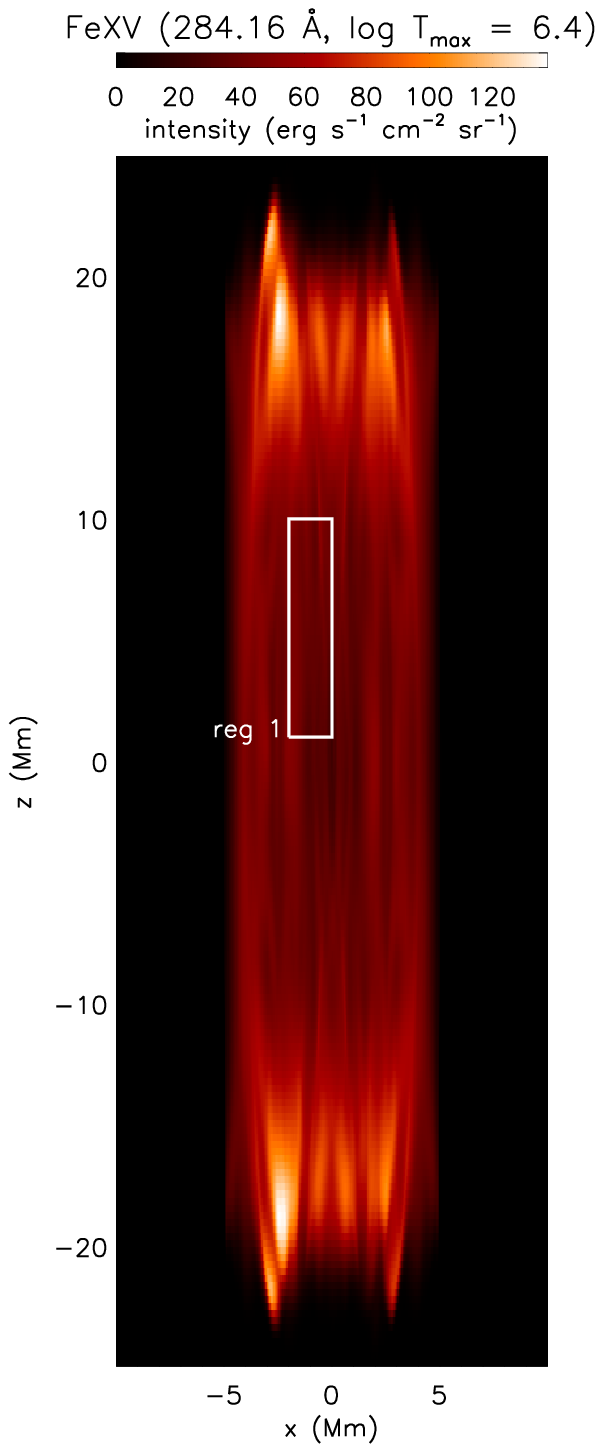}
			        \includegraphics[scale=0.45]{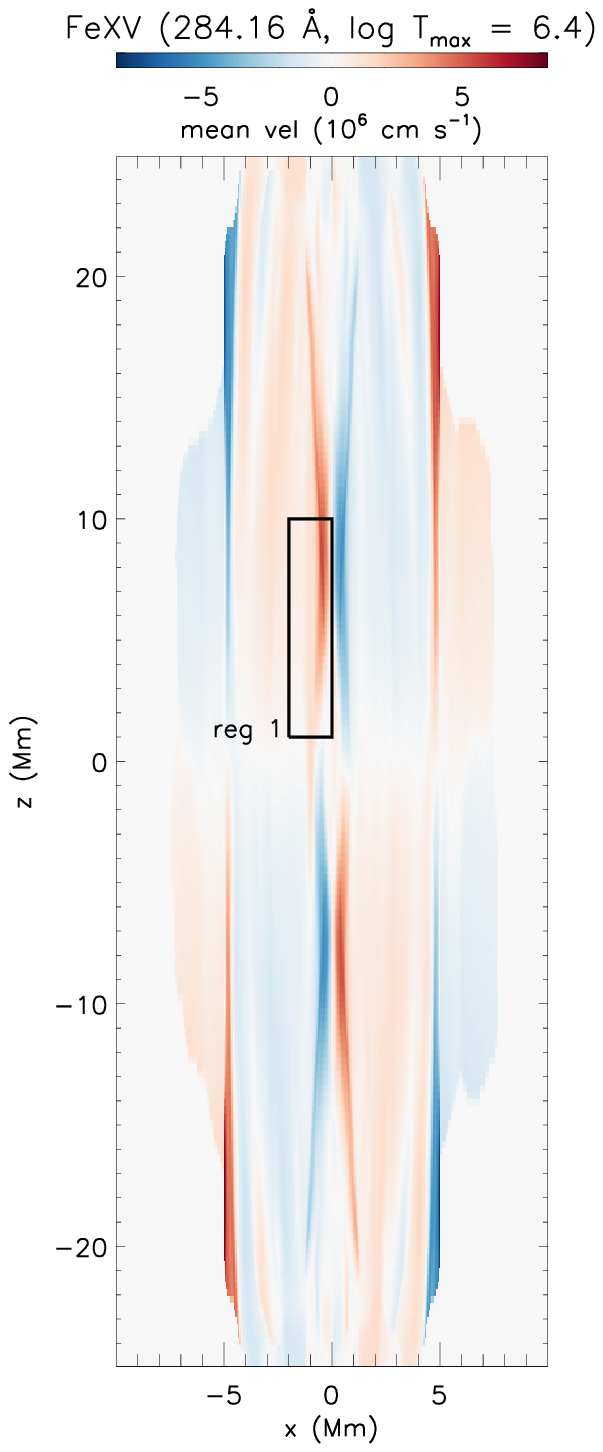}
			        \includegraphics[scale=0.45]{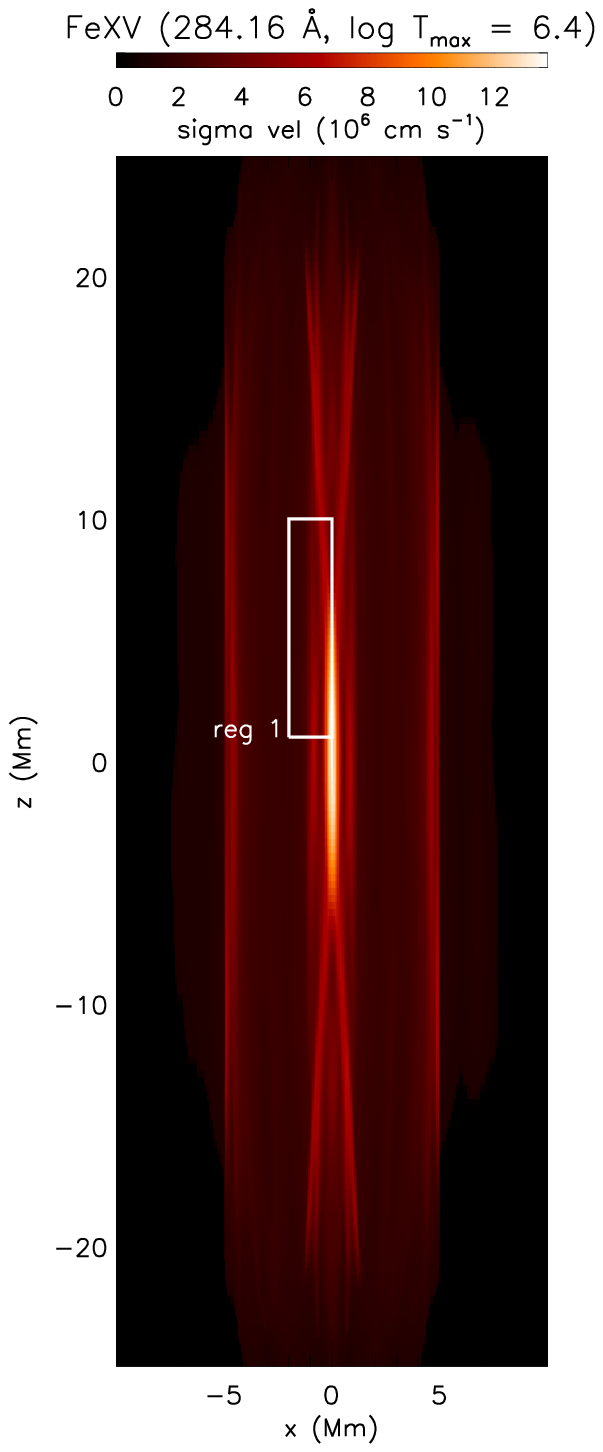}}
\resizebox{\hsize}{!}{\includegraphics[scale=0.45]{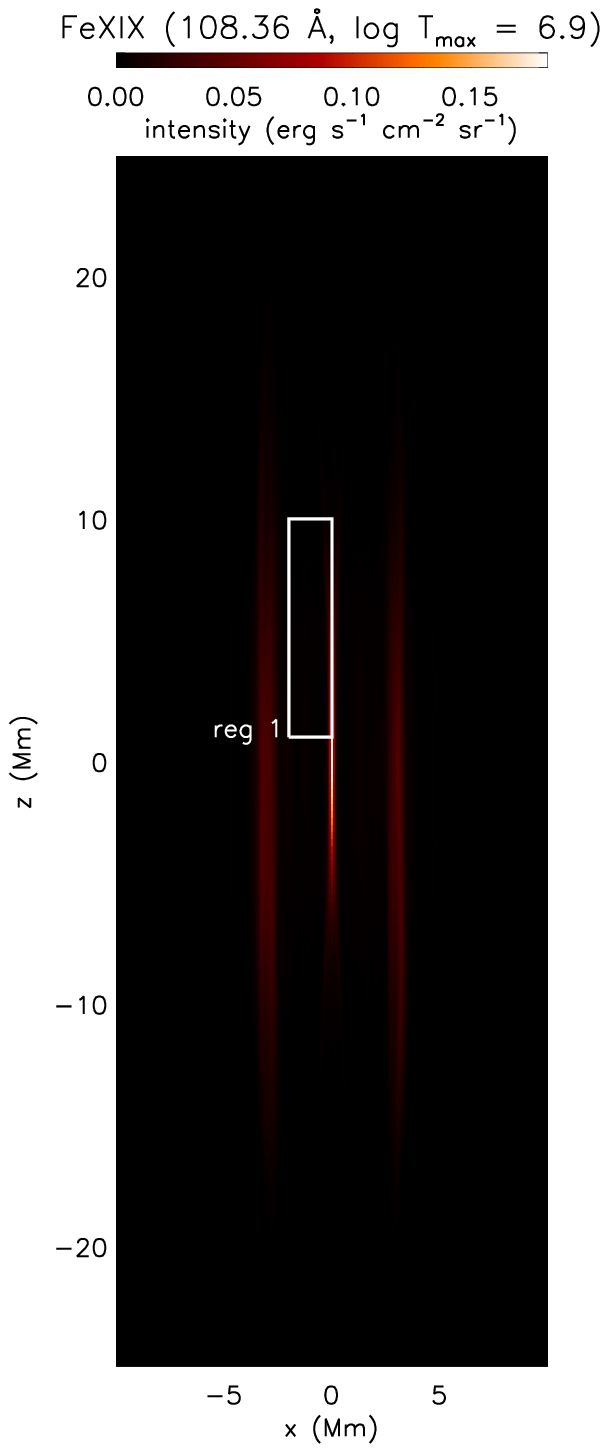}
			        \includegraphics[scale=0.45]{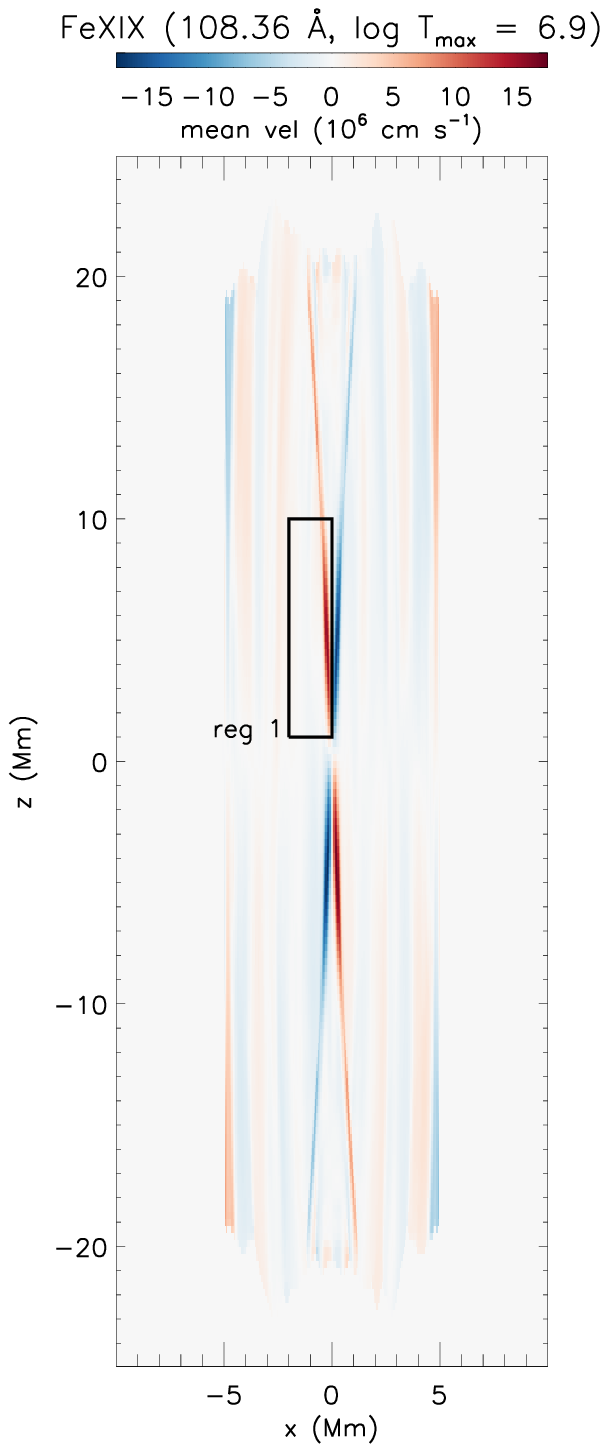}
			        \includegraphics[scale=0.45]{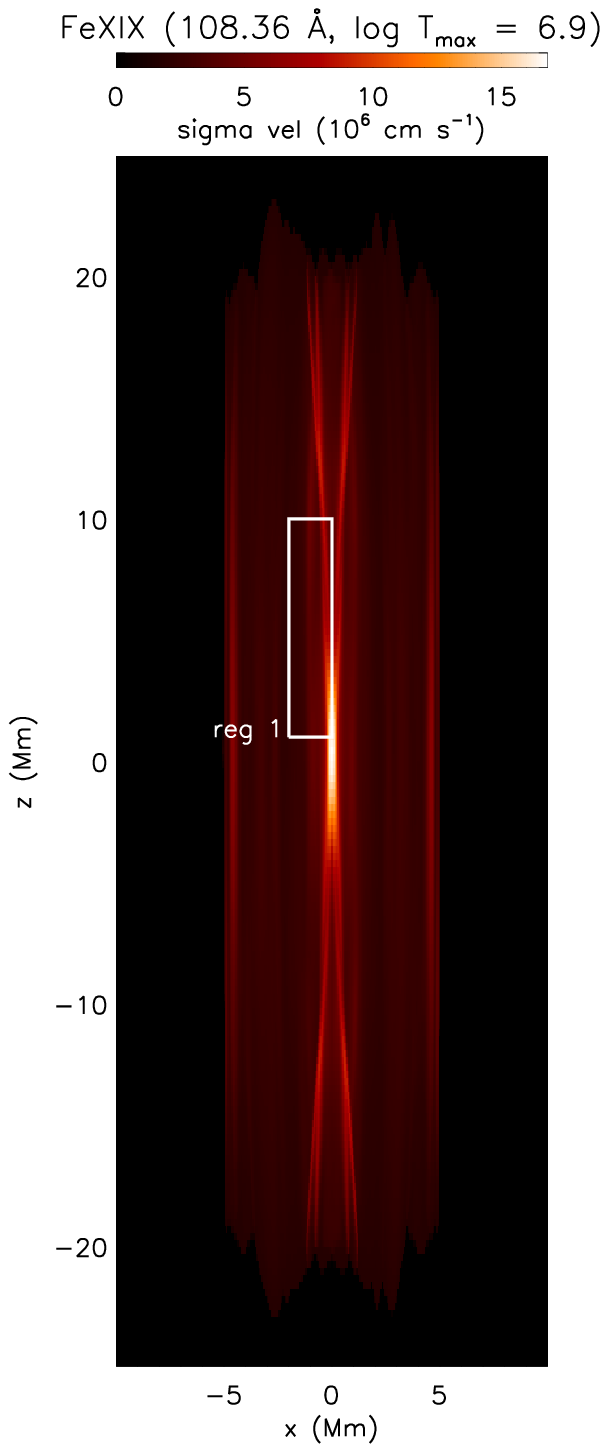}}
\caption{Maps of line intensity (left panels), mean velocity (central panels), and line width for the Fe\,XV line (top panels) and for the Fe\,XIX line (bottom panels), obtained from the $j_c=250$ simulation at $t=228$~s, and by integrating along the $y$ direction. The rectangular region indicates the region adopted for integrating line profiles and emission measure distribution displayed in Fig.~\ref{fig:linesmuse2}.}
\label{fig:linesmuse1}
\end{figure}

For each line we compute maps of intensity, shift, and width, assuming the simulation is observed along $y$, the jet propagation direction. The intensity map integrates line emission along the line of sight ($y$). Velocity and width maps account for thermal broadening and line shift, integrating along $y$ and computing mean velocity and width from the line profile in each map bin. The resulting emission maps for the $j_c=250$ simulation at $t=384~s$ are shown in Fig.~\ref{fig:linesmuse1}.

For Fe XV, emission at the jet time is mostly concentrated at the flux-tube footpoints, but significant brightness is also found throughout, even at the loop apex (domain centre) where density is lower.
The Doppler shift shows an interesting pattern: nearly zero near the domain centre, where two counterpropagating jets combine to cancel any significant shift.
In contrast, regions $\sim5~Mm$ further out (left and right of centre) show Doppler velocities of about $50~km/s$ — smaller than the jet peak velocity of $\sim250~km/s$, but large enough to stand out.
These regions form because the slingshot plasma runs along both flux tubes, whose footpoints were displaced in opposite directions; the outflows overlap only at the centre, where the jet cores align, and lie on different lines of sight elsewhere.
The line width is more pronounced at the domain centre, where the overlapping jet cores create bidirectional flows and high temperature, both widening the line profile.

The Fe XIX synthesis differs, given the different temperature it probes. 
This line is significantly strong only at the domain centre, where hot plasma is present; the footpoints and rest of the structure are very faint.
The Doppler velocity pattern is similar to Fe XV but narrower, reaching speeds up to $150~km/s$ — sensibly faster plasma than visible in Fe XV.
At the centre, where the line is more visible and spread (again due to bidirectional motion and high temperature), the line width is of order $100~km/s$.

\begin{figure}
\centering
\includegraphics[scale=0.70]{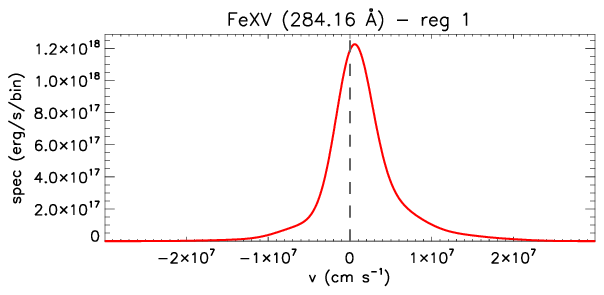}

\includegraphics[scale=0.70]{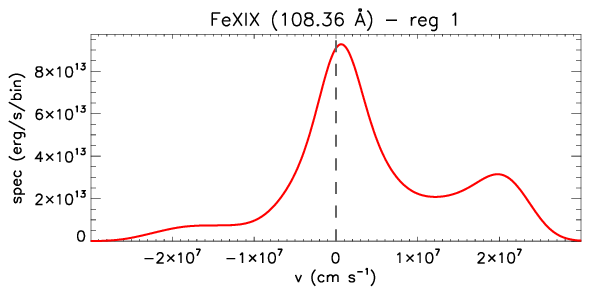}

\caption{Line profiles obtained by integrating the emerging spectra over a region that covers near the centre of the domain, with $x$ ranging from $-2~Mm$ to $0~Mm$, and $y$ ranging from $1~Mm$ and $10~Mm$ (see Fig.\ref{fig:linesmuse1}). The dimension of the velocity bin is 10$^5$~cm\,s$^{-1}$.}
\label{fig:linesmuse2}
\end{figure}

We investigate the diagnostic power of the selected line profiles further by spatially integrating the emerging spectra maps over a rectangular region near the domain centre, with $x$ from $-2~Mm$ to $0$ and $y$ from $0.1~Mm$ to $1~Mm$.
This region reasonably encompasses the high-temperature plasma heated during reconnection and the jet dynamics, though the field of view is much larger than the spatial resolution of modern instruments. This analysis should thus be viewed not as reconstructing an actual observation, but as a rough approximation of what observational signatures might remain from a reconnection event that is intrinsically local and fast.
The resulting spectra are shown in Fig.\ref{fig:linesmuse2}.
The Fe XV line remains much brighter than Fe XIX, by several orders of magnitude — a key indication of which signatures are more easily detectable.
This is a relatively common feature of coronal emission lines, since Fe XV's formation temperature is typical of the corona's thermal structure, whereas only plasma heated during transients reaches Fe XIX formation temperatures.
Despite the large region considered, the Fe XV line profile still shows a significant displacement of about $5~km/s$.
This small Doppler shift could signal much larger localised, undetected velocities, but a velocity of $\sim5~km/s$ is otherwise ordinary and would be difficult to detect in a coronal field of view full of similarly moving structures.
The Fe XIX line, in contrast, gives more definitive indications of jet plasma displacement, with an obvious additional peak $\sim200~km/s$ from the line centre. This signature is more directly linked to the jet, and stands out clearly for three reasons:
i) the field of view is offset from the domain centre, where bidirectional jets tend to cancel each other's contribution to the line shift;
ii) the jet plasma is hot and thus prominent in the Fe XIX line;
iii) the jet's substantial speed shifts its contribution away from the main peak, generating two distinctly measurable components.

\section{Discussion and conclusions}
\label{sec:conclusions}

In this paper, we extend a theoretical and modelling effort to describe the physical mechanisms and phenomena stemming from magnetic reconnection events in the solar corona.
Only recently has the detection of nanojets, i.e. post-reconnection outflows, shown a viable path to advance our analysis of magnetic reconnection and heating in the solar corona. This work thus aims to provide a wider and more accurate theoretical understanding of nanojet and nanoflare mechanisms to enable new detection capabilities.
This is key to validating the nanoflare theory as the main contributor of thermal energy in the solar corona. Indeed, several works have recently singled out magnetic reconnection events through nanojets \citep{2022ApJ...934..190S, 2022ApJ...938..122P, 2024A&A...692A.119C, 2025MNRAS.544.1758W}, though only in a number of favourable cases.

Specifically, we identify an important trait for the evolution of magnetic strands in the solar corona that facilitates reconnection in our setup. Reconnection occurs here because the initial condition is an unstable balance of pressure forces, with the two flux tubes symmetrically placed and their expansion quenched. As soon as the flux tubes move out of this configuration, the curvature of the magnetic field, and thus its tension, invariably leads to reconnection at the centre.
In this low-$\beta$ regime, the arrangement of magnetic tension is key to forcing reconnection against the outward pressure forces generated by the approaching reconnecting flux.
In other words, transverse MHD waves appear to be an unavoidable effect paired with the mechanisms unfolding before and after reconnection.
Observations in \citet{2022ApJ...934..190S} also point to instability development as a key ingredient for the system's evolution towards reconnection. Additionally, \citet{2025ApJ...995...94C} identifies the unstable unwinding of a magnetic flux rope prior to reconnection outflow.
Other models, even using different mechanisms to drive the braiding of magnetic threads, in our opinion agree with this scenario. Several works explicitly consider the kink instability as a possible mechanism to eventually force reconnection \citep{2023A&A...678A..40C, 2023MNRAS.518.1584R, 2024A&A...689A.184C, 2025A&A...695A..40C, 2025ApJ...995L..65R}.
In \citet{2025A&A...699A.106S}, reconnection becomes the system's runaway once two plasmoids form and are forced to merge by their dynamics; in \citet{2024MNRAS.530.2361B}, reconnection outflows are identified as soon as turbulent motion forces field lines to reconnect.
Observations \citep{2025ApJ...988L..65B, 2025A&A...702A.189T} show that reconnection outflows in flaring and dynamic regions have higher speeds and release more energy than those in quasi-statically evolving regions.

We carry out our investigation by changing the critical current above which magnetic resistivity switches on, to study how heating and velocity evolve as more free magnetic energy accumulates. While this remains an open question, other studies have already shown that the actual value of magnetic resistivity is not crucial in determining reconnection-induced heating in our regime \citep{1996JGR...10113445G, 2012ApJ...747..109N}.

We also explore how weaker or stronger jets result from allowing more or less magnetic stress to accumulate, by tuning the parameter controlling the onset of magnetic resistivity.
This experiment yields some interesting results. 
Dissipated magnetic energy is not strictly proportional to accumulated magnetic energy, mostly because a lower critical current allows more frequent, though more modest, ohmic heating, making the overall heating smaller than for less frequent but more energetic events.
For temperature, an initial trend correlates higher temperatures with more accumulated magnetic stress, but it soon plateaus around $7~MK$, in good agreement with \citep{2025ApJ...995L..65R}.
Similarly, jet speed initially grows, but once simulations reach speeds of about $350~km/s$, none of the critical currents used produce higher velocities.
This leads to ambiguous dynamics when trying to use jet properties as a proxy for the heating or temperature of the corresponding nanoflare.
This remains an outstanding challenge for modelling and observations, as several factors, inherent to the reconnection or its environment, seem to determine the slingshot speed or maximum temperature reached. Observationally, only a few studies \citep[e.g.][]{2021NatAs...5...54A,2025ApJ...982..147M} have independently addressed nanojet velocity and nanoflare temperature, while \citet{2025ApJ...995L..65R} recently showed that the ambient magnetic field plays a crucial role in available heating, though not necessarily for nanojet velocity.

Additional simulations in our models also address whether a wider energy range of reconnection events can be described with this approach.
We find that reconnecting flux tubes with smaller cross section still generate outflows at $\sim100~km/s$, but with much smaller overall heating, down to almost $10^{24}~erg$, much closer to the nanoflare energy range.
Smaller cross-section flux tubes can correspond to intraloop structures, describing a scenario where several such events cohabit and occur nearly simultaneously in a coronal loop.
At the same time, \citet{2025MNRAS.544.1758W}, \citet{2025A&A...702A.189T}, and \citet{2025ApJ...985L..12G} have recently observed reconnection events up to $10^{27}~erg$ in larger coronal structures with very large outflow speeds, up to $\sim500~km/s$. This suggests the size of the involved flux tubes or strands is a key parameter determining energy release.
Future work will need to address, in similar or more realistic scenarios, how nanojet and nanoflare dynamics and energetics are determined by the free magnetic energy injected into the coronal loop from photospheric drivers, rather than how efficient is its dissipation as addressed here. Several works have already tried to measure such energy inputs, and it would be useful to link them to models \citep{2015PASJ...67...18W, 2023ApJ...956L...1C}.
It should be noted, however, that the kinetic energy, i.e.,the nanojet, remains a small fraction of the available energy, as in our simulations and as already shown in previous works \citep[e.g.][]{2007ApJ...657L..47R, 2008ApJ...677.1348R}.

Finally, we have synthesised the emission from the plasma in these simulations, focusing on the Fe XV and Fe XIX lines, key to understanding temperature evolution during the impulsive heating from reconnection and the associated outflows.
We find that detection of Fe XIX can reveal the occurrence of a reconnection event, as its emission is simultaneous with the sudden plasma heating.
The Doppler shift of both lines can also be used to find nanojet motions. However, several effects (line-of-sight superposition of structures and the bidirectional nature of the nanojets) can significantly reduce the measured velocity. 
This means proper nanojets travelling at $\sim100~km/s$ can appear as $\sim10~km/s$ or smaller components.
This poses a significant observational challenge, as it is difficult to know when such a modest measured velocity signals fast reconnection outflows.
This can partly explain why the nature of these outflows has remained so elusive.
Similar conclusions on line shift and broadening were found in \citep{2023A&A...678A..40C, 2024A&A...689A.184C, 2025A&A...695A..40C}, where reconnection is triggered by kink instability and MHD avalanche \citep{2016ApJ...817....5H}, supporting the multifaceted nature of nanojets and the many ways reconnection can originate in the solar corona.

Similarly, \citet{2025ApJ...994..139J} and \citet{2026MNRAS.545f2180B} have developed realistic coronal loop models where braiding of field lines is fostered through boundary conditions mimicking photospheric motions more turbulent and chaotic than ours. Our simulations, instead, aim to describe a simpler configuration to get a deeper physical insight into the properties of the heating and reconnection outflows in a system where such effects can be isolated more easily.

Crucially, our model, while realistic, is stripped of the complexity of an actual heterogeneous and dynamic coronal environment such as an active region.
This has allowed us to focus on the fundamental properties of nanojets and nanoflares, illustrating some of the underlying physics.

This work suggests several ways forward for studying coronal heating due to magnetic reconnection, many of which will become investigable with the upcoming NASA mission MUSE, offering the spatial and time resolution as well as spectroscopic capabilities to tackle most of the challenges illustrated here.

\begin{acknowledgements}
      \\
      This work made use of the HPC system MEUSA, part of the Sistema Computazionale per l'Astrofisica Numerica (SCAN) of INAF-Osservatorio Astronomico di Palermo.
      We acknowledge ISCRA for awarding this project access to the LEONARDO supercomputer, owned by the EuroHPC Joint Undertaking, hosted by CINECA (Italy).
      P.P. and F.R. acknowledge support from ASI/INAF agreement no. 2022-29-HH.0. G.C. and P.T. were supported by contract 4105785828 (MUSE) to the Smithsonian Astrophysical Observatory, and by NASA grant 80NSSC21K1684.

\end{acknowledgements}

%
\bibliographystyle{aa} 
%

\bibliography{ref}

@string{apj="ApJ\ "}

@string{apjs="ApJS\ "}

@string{mnras="MNRAS\ "}

@string{pasj={Publ. Astr. Soc. Japan}}

@string{ssr="Space Sci. Rev.\ "}

@ARTICLE{Rosner1978,
   author = {{Rosner}, R. and {Tucker}, W.~H. and {Vaiana}, G.~S.},
    title = "{Dynamics of the quiescent solar corona}",
  journal = {\apj},
     year = 1978,
    month = mar,
   volume = 220,
    pages = {643-645},
      doi = {10.1086/155949},
   adsurl = {http://adsabs.harvard.edu/cgi-bin/nph-bib_query?bibcode=1978ApJ...220..643R&db_key=AST}
}

@ARTICLE{ParnellDeMoortel2012,
   author = {{Parnell}, C.~E. and {De Moortel}, I.},
    title = "{A contemporary view of coronal heating}",
  journal = {Philosophical Transactions of the Royal Society of London Series A},
archivePrefix = "arXiv",
   eprint = {1206.6097},
 primaryClass = "astro-ph.SR",
     year = 2012,
    month = jul,
   volume = 370,
    pages = {3217-3240},
      doi = {10.1098/rsta.2012.0113},
   adsurl = {http://adsabs.harvard.edu/abs/2012RSPTA.370.3217P}
}

@ARTICLE{DeMoortelBrowning2015,
   author = {{De Moortel}, I. and {Browning}, P.},
    title = "{Recent advances in coronal heating}",
  journal = {Philosophical Transactions of the Royal Society of London Series A},
archivePrefix = "arXiv",
   eprint = {1510.00977},
 primaryClass = "astro-ph.SR",
     year = 2015,
    month = apr,
   volume = 373,
    pages = {20140269-20140269},
      doi = {10.1098/rsta.2014.0269},
   adsurl = {http://adsabs.harvard.edu/abs/2015RSPTA.37340269D}
}

@ARTICLE{Reale2016,
   author = {{Reale}, F. and {Orlando}, S. and {Guarrasi}, M. and {Mignone}, A. and 
	{Peres}, G. and {Hood}, A.~W. and {Priest}, E.~R.},
    title = "{3D MHD modeling of twisted coronal loops}",
  journal = {\apj},
archivePrefix = "arXiv",
   eprint = {1607.05500},
 primaryClass = "astro-ph.SR",
     year = 2016,
    month = oct,
   volume = 830,
      eid = {21},
    pages = {21},
      doi = {10.3847/0004-637X/830/1/21},
   adsurl = {http://adsabs.harvard.edu/abs/2016ApJ...830...21R}
}

@ARTICLE{Parker1988,
   author = {{Parker}, E.~N.},
    title = "{Nanoflares and the solar X-ray corona}",
  journal = {\apj},
     year = 1988,
    month = jul,
   volume = 330,
    pages = {474-479},
      doi = {10.1086/166485},
   adsurl = {http://adsabs.harvard.edu/abs/1988ApJ...330..474P}
}

@ARTICLE{Klimchuk2015,
   author = {{Klimchuk}, J.~A.},
    title = "{Key aspects of coronal heating}",
  journal = {Philosophical Transactions of the Royal Society of London Series A},
archivePrefix = "arXiv",
   eprint = {1410.5660},
 primaryClass = "astro-ph.SR",
     year = 2015,
    month = apr,
   volume = 373,
    pages = {20140256-20140256},
      doi = {10.1098/rsta.2014.0256},
   adsurl = {http://adsabs.harvard.edu/abs/2015RSPTA.37340256K}
}

@article{Mignone_2012ApJS..198....7M,
	Adsurl = {http://adsabs.harvard.edu/abs/2012ApJS..198....7M},
	Archiveprefix = {arXiv},
	Author = {{Mignone}, A. and {Zanni}, C. and {Tzeferacos}, P. and {van Straalen}, B. and {Colella}, P. and {Bodo}, G.},
	Doi = {10.1088/0067-0049/198/1/7},
	Eid = {7},
	Eprint = {1110.0740},
	Journal = {\apjs},
	Month = jan,
	Pages = {7},
	Primaryclass = {astro-ph.HE},
	Title = {{The PLUTO Code for Adaptive Mesh Computations in Astrophysical Fluid Dynamics}},
	Volume = 198,
	Year = 2012}

@ARTICLE{2021NatAs...5...54A,
       author = {{Antolin}, Patrick and {Pagano}, Paolo and {Testa}, Paola and {Petralia}, Antonino and {Reale}, Fabio},
        title = "{Reconnection nanojets in the solar corona}",
      journal = {Nature Astronomy},
         year = 2021,
        month = jan,
       volume = {5},
        pages = {54-62},
          doi = {10.1038/s41550-020-1199-8},
       adsurl = {https://ui.adsabs.harvard.edu/abs/2021NatAs...5...54A}
}

@ARTICLE{2014LRSP...11....4R,
       author = {{Reale}, Fabio},
        title = "{Coronal Loops: Observations and Modeling of Confined Plasma}",
      journal = {Living Reviews in Solar Physics},
         year = 2014,
        month = jul,
       volume = {11},
       number = {1},
          eid = {4},
        pages = {4},
          doi = {10.12942/lrsp-2014-4},
       adsurl = {https://ui.adsabs.harvard.edu/abs/2014LRSP...11....4R}
}

@ARTICLE{2021A&A...656A.141P,
       author = {{Pagano}, P. and {Antolin}, P. and {Petralia}, A.},
        title = "{Modelling of asymmetric nanojets in coronal loops}",
      journal = {\aap},
         year = 2021,
        month = dec,
       volume = {656},
          eid = {A141},
        pages = {A141},
          doi = {10.1051/0004-6361/202141030},
archivePrefix = {arXiv},
       eprint = {2109.04854},
 primaryClass = {astro-ph.SR},
       adsurl = {https://ui.adsabs.harvard.edu/abs/2021A&A...656A.141P}
}

@ARTICLE{2023Symm...15..627C,
       author = {{Cozzo}, Gabriele and {Pagano}, Paolo and {Petralia}, Antonino and {Reale}, Fabio},
        title = "{Asymmetric Twisting of Coronal Loops}",
      journal = {Symmetry},
         year = 2023,
        month = mar,
       volume = {15},
       number = {3},
          eid = {627},
        pages = {627},
          doi = {10.3390/sym15030627},
       adsurl = {https://ui.adsabs.harvard.edu/abs/2023Symm...15..627C}
}

@ARTICLE{2023A&A...678A..40C,
       author = {{Cozzo}, G. and {Reid}, J. and {Pagano}, P. and {Reale}, F. and {Hood}, A.~W.},
        title = "{Coronal energy release by MHD avalanches. Effects on a structured, active region, multi-threaded coronal loop}",
      journal = {\aap},
         year = 2023,
        month = oct,
       volume = {678},
          eid = {A40},
        pages = {A40},
          doi = {10.1051/0004-6361/202346689},
archivePrefix = {arXiv},
       eprint = {2306.06047},
 primaryClass = {astro-ph.SR},
       adsurl = {https://ui.adsabs.harvard.edu/abs/2023A&A...678A..40C}
}

@ARTICLE{1989GeCoA..53..197A,
       author = {{Anders}, E. and {Grevesse}, N.},
        title = "{Abundances of the elements: Meteoritic and solar}",
      journal = {\gca},
         year = 1989,
        month = jan,
       volume = {53},
       number = {1},
        pages = {197-214},
          doi = {10.1016/0016-7037(89)90286-X},
       adsurl = {https://ui.adsabs.harvard.edu/abs/1989GeCoA..53..197A}
}

@ARTICLE{2013ApJ...772...71L,
       author = {{Landi}, E. and {Reale}, F.},
        title = "{Prominence Plasma Diagnostics through Extreme-ultraviolet Absorption}",
      journal = {\apj},
         year = 2013,
        month = jul,
       volume = {772},
       number = {1},
          eid = {71},
        pages = {71},
          doi = {10.1088/0004-637X/772/1/71},
archivePrefix = {arXiv},
       eprint = {1209.2934},
 primaryClass = {astro-ph.SR},
       adsurl = {https://ui.adsabs.harvard.edu/abs/2013ApJ...772...71L}
}

@ARTICLE{2001JGR...10625165L,
       author = {{Linker}, J.~A. and {Lionello}, R. and {Miki{\'c}}, Z. and {Amari}, T.},
        title = "{Magnetohydrodynamic modeling of prominence formation within a helmet streamer}",
      journal = {\jgr},
         year = 2001,
        month = nov,
       volume = {106},
       number = {A11},
        pages = {25165-25176},
          doi = {10.1029/2000JA004020},
       adsurl = {https://ui.adsabs.harvard.edu/abs/2001JGR...10625165L}
}

@ARTICLE{2009ApJ...690..902L,
       author = {{Lionello}, Roberto and {Linker}, Jon A. and {Miki{\'c}}, Zoran},
        title = "{Multispectral Emission of the Sun During the First Whole Sun Month: Magnetohydrodynamic Simulations}",
      journal = {\apj},
         year = 2009,
        month = jan,
       volume = {690},
       number = {1},
        pages = {902-912},
          doi = {10.1088/0004-637X/690/1/902},
       adsurl = {https://ui.adsabs.harvard.edu/abs/2009ApJ...690..902L}
}

@ARTICLE{2013ApJ...773...94M,
       author = {{Miki{\'c}}, Zoran and {Lionello}, Roberto and {Mok}, Yung and {Linker}, Jon A. and {Winebarger}, Amy R.},
        title = "{The Importance of Geometric Effects in Coronal Loop Models}",
      journal = {\apj},
         year = 2013,
        month = aug,
       volume = {773},
       number = {2},
          eid = {94},
        pages = {94},
          doi = {10.1088/0004-637X/773/2/94},
       adsurl = {https://ui.adsabs.harvard.edu/abs/2013ApJ...773...94M}
}

@ARTICLE{1981ApJ...243..288S,
       author = {{Serio}, S. and {Peres}, G. and {Vaiana}, G.~S. and {Golub}, L. and {Rosner}, R.},
        title = "{Closed coronal structures. II - Generalized hydrostatic model}",
      journal = {\apj},
         year = 1981,
        month = jan,
       volume = {243},
        pages = {288-300},
          doi = {10.1086/158597},
       adsurl = {https://ui.adsabs.harvard.edu/abs/1981ApJ...243..288S}
}

@ARTICLE{2014A&A...564A..48G,
       author = {{Guarrasi}, M. and {Reale}, F. and {Orlando}, S. and {Mignone}, A. and {Klimchuk}, J.~A.},
        title = "{MHD modeling of coronal loops: the transition region throat}",
      journal = {\aap},
         year = 2014,
        month = apr,
       volume = {564},
          eid = {A48},
        pages = {A48},
          doi = {10.1051/0004-6361/201322848},
archivePrefix = {arXiv},
       eprint = {1402.0338},
 primaryClass = {astro-ph.SR},
       adsurl = {https://ui.adsabs.harvard.edu/abs/2014A&A...564A..48G}
}

@ARTICLE{2020ApJ...888....3D,
       author = {{De Pontieu}, Bart and {Mart{\'\i}nez-Sykora}, Juan and {Testa}, Paola and {Winebarger}, Amy R. and {Daw}, Adrian and {Hansteen}, Viggo and {Cheung}, Mark C.~M. and {Antolin}, Patrick},
        title = "{The Multi-slit Approach to Coronal Spectroscopy with the Multi-slit Solar Explorer (MUSE)}",
      journal = {\apj},
         year = 2020,
        month = jan,
       volume = {888},
       number = {1},
          eid = {3},
        pages = {3},
          doi = {10.3847/1538-4357/ab5b03},
archivePrefix = {arXiv},
       eprint = {1909.08818},
 primaryClass = {astro-ph.IM},
       adsurl = {https://ui.adsabs.harvard.edu/abs/2020ApJ...888....3D}
}

@ARTICLE{2022ApJ...934..190S,
       author = {{Sukarmadji}, A. Ramada C. and {Antolin}, Patrick and {McLaughlin}, James A.},
        title = "{Observations of Instability-driven Nanojets in Coronal Loops}",
      journal = {\apj},
         year = 2022,
        month = aug,
       volume = {934},
       number = {2},
          eid = {190},
        pages = {190},
          doi = {10.3847/1538-4357/ac7870},
archivePrefix = {arXiv},
       eprint = {2202.10960},
 primaryClass = {astro-ph.SR},
       adsurl = {https://ui.adsabs.harvard.edu/abs/2022ApJ...934..190S}
}

@ARTICLE{2022ApJ...938..122P,
       author = {{Patel}, Ritesh and {Pant}, Vaibhav},
        title = "{Hi-C 2.1 Observations of Reconnection Nanojets}",
      journal = {\apj},
         year = 2022,
        month = oct,
       volume = {938},
       number = {2},
          eid = {122},
        pages = {122},
          doi = {10.3847/1538-4357/ac92e5},
archivePrefix = {arXiv},
       eprint = {2209.06449},
 primaryClass = {astro-ph.SR},
       adsurl = {https://ui.adsabs.harvard.edu/abs/2022ApJ...938..122P}
}

@ARTICLE{2025MNRAS.544.1758W,
       author = {{Wallace}, Tarhik and {Antolin}, Patrick},
        title = "{Reconnection nanojets associated with a prominence eruption observed with Solar Orbiter/EUI-HRI}",
      journal = {\mnras},
         year = 2025,
        month = dec,
       volume = {544},
       number = {2},
        pages = {1758-1768},
          doi = {10.1093/mnras/staf1879},
archivePrefix = {arXiv},
       eprint = {2510.25068},
 primaryClass = {astro-ph.SR},
       adsurl = {https://ui.adsabs.harvard.edu/abs/2025MNRAS.544.1758W}
}

@ARTICLE{2025ApJ...995...94C,
       author = {{Chen}, Hechao and {Tian}, Hui and {Priest}, Eric R. and {Prior}, Christopher B. and {Xia}, Chun and {Yeates}, Anthony R. and {Yan}, Xiaoli and {Duan}, Yadan and {Hou}, Zhenyong and {Huang}, Zhenghua and {Rice}, Oliver E.~K.},
        title = "{Witnessing Magnetic Reconnection in Tangled Superpenumbral Fibrils around a Sunspot}",
      journal = {\apj},
         year = 2025,
        month = dec,
       volume = {995},
       number = {1},
          eid = {94},
        pages = {94},
          doi = {10.3847/1538-4357/ae12e9},
archivePrefix = {arXiv},
       eprint = {2509.23636},
 primaryClass = {astro-ph.SR},
       adsurl = {https://ui.adsabs.harvard.edu/abs/2025ApJ...995...94C}
}

@ARTICLE{2025A&A...702A.189T,
       author = {{Tan}, Song and {Warmuth}, Alexander and {Schuller}, Fr{\'e}d{\'e}ric and {Shen}, Yuandeng and {Mitchell}, Jake A.~J. and {Shi}, Fanpeng},
        title = "{Extremely diverse coronal jets accompanying an erupting filament captured by Solar Orbiter}",
      journal = {\aap},
         year = 2025,
        month = oct,
       volume = {702},
          eid = {A189},
        pages = {A189},
          doi = {10.1051/0004-6361/202555297},
archivePrefix = {arXiv},
       eprint = {2509.04741},
 primaryClass = {astro-ph.SR},
       adsurl = {https://ui.adsabs.harvard.edu/abs/2025A&A...702A.189T}
}

@ARTICLE{2025ApJ...988L..65B,
       author = {{Bura}, Annu and {Shrivastav}, Arpit Kumar and {Patel}, Ritesh and {Samanta}, Tanmoy and {Nayak}, Sushree S. and {Ghosh}, Ananya and {Sow Mondal}, Shanwlee and {Pant}, Vaibhav and {Seaton}, Daniel B.},
        title = "{Dynamics of Reconnection Nanojets in Eruptive and Confined Solar Flares}",
      journal = {\apjl},
         year = 2025,
        month = aug,
       volume = {988},
       number = {2},
          eid = {L65},
        pages = {L65},
          doi = {10.3847/2041-8213/adef0f},
archivePrefix = {arXiv},
       eprint = {2507.04639},
 primaryClass = {astro-ph.SR},
       adsurl = {https://ui.adsabs.harvard.edu/abs/2025ApJ...988L..65B}
}

@ARTICLE{2025ApJ...985L..12G,
       author = {{Gao}, Yuhang and {Tian}, Hui and {Berghmans}, David and {Duan}, Yadan and {Van Doorsselaere}, Tom and {Chen}, Hechao and {Kraaikamp}, Emil},
        title = "{Reconnection Nanojets in an Erupting Solar Filament with Unprecedented High Speeds}",
      journal = {\apjl},
         year = 2025,
        month = may,
       volume = {985},
       number = {1},
          eid = {L12},
        pages = {L12},
          doi = {10.3847/2041-8213/add33a},
archivePrefix = {arXiv},
       eprint = {2504.20663},
 primaryClass = {astro-ph.SR},
       adsurl = {https://ui.adsabs.harvard.edu/abs/2025ApJ...985L..12G}
}

@ARTICLE{2025ApJ...982..147M,
       author = {{Mishra}, Sudheer K. and {Srivastava}, A.~K. and {Rajaguru}, S.~P. and {Jel{\'\i}nek}, P.},
        title = "{Formation of Jet-driven Forced Reconnection Region and Associated Plasma Blobs in a Prominence Segment}",
      journal = {\apj},
         year = 2025,
        month = apr,
       volume = {982},
       number = {2},
          eid = {147},
        pages = {147},
          doi = {10.3847/1538-4357/adb8db},
archivePrefix = {arXiv},
       eprint = {2502.12889},
 primaryClass = {astro-ph.SR},
       adsurl = {https://ui.adsabs.harvard.edu/abs/2025ApJ...982..147M}
}

@ARTICLE{2024A&A...692A.119C,
       author = {{Chen}, Yajie and {Mandal}, Sudip and {Peter}, Hardi and {Chitta}, Lakshmi Pradeep},
        title = "{Bidirectional propagating brightenings in arch filament systems observed by Solar Orbiter/EUI}",
      journal = {\aap},
         year = 2024,
        month = dec,
       volume = {692},
          eid = {A119},
        pages = {A119},
          doi = {10.1051/0004-6361/202451069},
archivePrefix = {arXiv},
       eprint = {2411.07876},
 primaryClass = {astro-ph.SR},
       adsurl = {https://ui.adsabs.harvard.edu/abs/2024A&A...692A.119C}
}

@ARTICLE{2025ApJ...995L..65R,
       author = {{Reale}, F. and {Cozzo}, G. and {Pagano}, P. and {Testa}, P.},
        title = "{On the Connection of Coronal Loop Plasma with the Ambient Magnetic Field}",
      journal = {\apjl},
         year = 2025,
        month = dec,
       volume = {995},
       number = {2},
          eid = {L65},
        pages = {L65},
          doi = {10.3847/2041-8213/ae2914},
       adsurl = {https://ui.adsabs.harvard.edu/abs/2025ApJ...995L..65R}
}

@ARTICLE{2025A&A...695A..40C,
       author = {{Cozzo}, G. and {Pagano}, P. and {Reale}, F. and {Testa}, P. and {Petralia}, A. and {Martinez-Sykora}, J. and {Hansteen}, V. and {De Pontieu}, B.},
        title = "{Coronal energy release by MHD avalanches: III. Identification of a reconnection outflow from a nanoflare}",
      journal = {\aap},
         year = 2025,
        month = mar,
       volume = {695},
          eid = {A40},
        pages = {A40},
          doi = {10.1051/0004-6361/202452426},
archivePrefix = {arXiv},
       eprint = {2502.01796},
 primaryClass = {astro-ph.SR},
       adsurl = {https://ui.adsabs.harvard.edu/abs/2025A&A...695A..40C}
}

@ARTICLE{2024A&A...689A.184C,
       author = {{Cozzo}, G. and {Reid}, J. and {Pagano}, P. and {Reale}, F. and {Testa}, P. and {Hood}, A.~W. and {Argiroffi}, C. and {Petralia}, A. and {Alaimo}, E. and {D'Anca}, F. and {Sciortino}, L. and {Todaro}, M. and {Lo Cicero}, U. and {Barbera}, M. and {de Pontieu}, B. and {Martinez-Sykora}, J.},
        title = "{Coronal energy release by MHD avalanches: II. EUV line emission from a multi-threaded coronal loop}",
      journal = {\aap},
         year = 2024,
        month = sep,
       volume = {689},
          eid = {A184},
        pages = {A184},
          doi = {10.1051/0004-6361/202450644},
archivePrefix = {arXiv},
       eprint = {2406.11701},
 primaryClass = {astro-ph.SR},
       adsurl = {https://ui.adsabs.harvard.edu/abs/2024A&A...689A.184C}
}

@ARTICLE{2016ApJ...817....5H,
       author = {{Hood}, A.~W. and {Cargill}, P.~J. and {Browning}, P.~K. and {Tam}, K.~V.},
        title = "{An MHD Avalanche in a Multi-threaded Coronal Loop.}",
      journal = {\apj},
         year = 2016,
        month = jan,
       volume = {817},
       number = {1},
          eid = {5},
        pages = {5},
          doi = {10.3847/0004-637X/817/1/5},
archivePrefix = {arXiv},
       eprint = {1512.00628},
 primaryClass = {astro-ph.SR},
       adsurl = {https://ui.adsabs.harvard.edu/abs/2016ApJ...817....5H}
}

@ARTICLE{2023MNRAS.518.1584R,
       author = {{Reid}, J. and {Threlfall}, J. and {Hood}, A.~W.},
        title = "{Self-consistent nanoflare heating in model active regions: MHD avalanches}",
      journal = {\mnras},
         year = 2023,
        month = jan,
       volume = {518},
       number = {1},
        pages = {1584-1600},
          doi = {10.1093/mnras/stac3188},
       adsurl = {https://ui.adsabs.harvard.edu/abs/2023MNRAS.518.1584R}
}

@ARTICLE{2024MNRAS.530.2361B,
       author = {{Breu}, C.~A. and {Peter}, H. and {Solanki}, S.~K. and {Cameron}, R. and {De Moortel}, I.},
        title = "{Non-thermal broadening of coronal lines in a 3D MHD loop model}",
      journal = {\mnras},
         year = 2024,
        month = may,
       volume = {530},
       number = {2},
        pages = {2361-2377},
          doi = {10.1093/mnras/stae899},
archivePrefix = {arXiv},
       eprint = {2404.00127},
 primaryClass = {astro-ph.SR},
       adsurl = {https://ui.adsabs.harvard.edu/abs/2024MNRAS.530.2361B}
}

@ARTICLE{2025A&A...699A.106S,
       author = {{Sen}, Samrat and {Moreno-Insertis}, Fernando},
        title = "{Merging plasmoids and nanojet-like ejections in a coronal current sheet}",
      journal = {\aap},
         year = 2025,
        month = jul,
       volume = {699},
          eid = {A106},
        pages = {A106},
          doi = {10.1051/0004-6361/202453595},
archivePrefix = {arXiv},
       eprint = {2505.02733},
 primaryClass = {astro-ph.SR},
       adsurl = {https://ui.adsabs.harvard.edu/abs/2025A&A...699A.106S}
}

@ARTICLE{2020SSRv..216..140V,
       author = {{Van Doorsselaere}, Tom and {Srivastava}, Abhishek K. and {Antolin}, Patrick and {Magyar}, Norbert and {Vasheghani Farahani}, Soheil and {Tian}, Hui and {Kolotkov}, Dmitrii and {Ofman}, Leon and {Guo}, Mingzhe and {Arregui}, I{\~n}igo and {De Moortel}, Ineke and {Pascoe}, David},
        title = "{Coronal Heating by MHD Waves}",
      journal = {\ssr},
         year = 2020,
        month = dec,
       volume = {216},
       number = {8},
          eid = {140},
        pages = {140},
          doi = {10.1007/s11214-020-00770-y},
archivePrefix = {arXiv},
       eprint = {2012.01371},
 primaryClass = {astro-ph.SR},
       adsurl = {https://ui.adsabs.harvard.edu/abs/2020SSRv..216..140V}
}

@ARTICLE{2022FrASS...920116A,
       author = {{Antolin}, Patrick and {Froment}, Clara},
        title = "{Multi-Scale Variability of Coronal Loops Set by Thermal Non-Equilibrium and Instability as a Probe for Coronal Heating}",
      journal = {Frontiers in Astronomy and Space Sciences},
         year = 2022,
        month = mar,
       volume = {9},
          eid = {820116},
        pages = {820116},
          doi = {10.3389/fspas.2022.820116},
       adsurl = {https://ui.adsabs.harvard.edu/abs/2022FrASS...920116A}
}

@ARTICLE{2023ApJ...957...25J,
       author = {{Judge}, P.~G.},
        title = "{Steadiness of Coronal Heating}",
      journal = {\apj},
         year = 2023,
        month = nov,
       volume = {957},
       number = {1},
          eid = {25},
        pages = {25},
          doi = {10.3847/1538-4357/acf83a},
archivePrefix = {arXiv},
       eprint = {2309.05164},
 primaryClass = {astro-ph.SR},
       adsurl = {https://ui.adsabs.harvard.edu/abs/2023ApJ...957...25J}
}

@ARTICLE{2014LRSP...11....1P,
       author = {{Parenti}, Susanna},
        title = "{Solar Prominences: Observations}",
      journal = {Living Reviews in Solar Physics},
         year = 2014,
        month = dec,
       volume = {11},
       number = {1},
          eid = {1},
        pages = {1},
          doi = {10.12942/lrsp-2014-1},
       adsurl = {https://ui.adsabs.harvard.edu/abs/2014LRSP...11....1P}
}

@ARTICLE{2022NatAs...6..942J,
       author = {{Jenkins}, Jack M. and {Keppens}, Rony},
        title = "{Resolving the solar prominence/filament paradox using the magnetic Rayleigh-Taylor instability}",
      journal = {Nature Astronomy},
         year = 2022,
        month = jul,
       volume = {6},
        pages = {942-950},
          doi = {10.1038/s41550-022-01705-z},
       adsurl = {https://ui.adsabs.harvard.edu/abs/2022NatAs...6..942J}
}

@ARTICLE{2022ApJ...926...52D,
       author = {{De Pontieu}, Bart and {Testa}, Paola and {Mart{\'\i}nez-Sykora}, Juan and {Antolin}, Patrick and {Karampelas}, Konstantinos and {Hansteen}, Viggo and {Rempel}, Matthias and {Cheung}, Mark C.~M. and {Reale}, Fabio and {Danilovic}, Sanja and {Pagano}, Paolo and {Polito}, Vanessa and {De Moortel}, Ineke and {N{\'o}brega-Siverio}, Daniel and {Van Doorsselaere}, Tom and {Petralia}, Antonino and {Asgari-Targhi}, Mahboubeh and {Boerner}, Paul and {Carlsson}, Mats and {Chintzoglou}, Georgios and {Daw}, Adrian and {DeLuca}, Edward and {Golub}, Leon and {Matsumoto}, Takuma and {Ugarte-Urra}, Ignacio and {McIntosh}, Scott W. and {the MUSE Team}},
        title = "{Probing the Physics of the Solar Atmosphere with the Multi-slit Solar Explorer (MUSE). I. Coronal Heating}",
      journal = {\apj},
         year = 2022,
        month = feb,
       volume = {926},
       number = {1},
          eid = {52},
        pages = {52},
          doi = {10.3847/1538-4357/ac4222},
archivePrefix = {arXiv},
       eprint = {2106.15584},
 primaryClass = {astro-ph.SR},
       adsurl = {https://ui.adsabs.harvard.edu/abs/2022ApJ...926...52D}
}

@ARTICLE{2026ApJ...998...75C,
       author = {{Cozzo}, G. and {Testa}, Paola and {Martinez-Sykora}, J. and {Reale}, F. and {Pagano}, P. and {Rappazzo}, F. and {Hansteen}, V. and {De Pontieu}, B. and {Petralia}, A. and {Alaimo}, E. and {Fiorentino}, F. and {D'Anca}, F. and {Sciortino}, L. and {Todaro}, M. and {Lo Cicero}, U. and {Barbera}, M.},
        title = "{3D MHD Simulations of Coronal Loops Heated via Magnetic Braiding. I. Continuous Driving}",
      journal = {\apj},
         year = 2026,
        month = feb,
       volume = {998},
       number = {1},
          eid = {75},
        pages = {75},
          doi = {10.3847/1538-4357/ae2d56},
archivePrefix = {arXiv},
       eprint = {2511.08726},
 primaryClass = {astro-ph.SR},
       adsurl = {https://ui.adsabs.harvard.edu/abs/2026ApJ...998...75C}
}

@ARTICLE{2026ApJ...998...76C,
       author = {{Cozzo}, G. and {Testa}, Paola and {Martinez-Sykora}, J. and {Pagano}, P. and {Reale}, F. and {Rappazzo}, F. and {Hansteen}, V. and {De Pontieu}, B.},
        title = "{3D MHD Simulations of Coronal Loops Heated via Magnetic Braiding. II. Automatic Detection of Reconnection Outflows and Statistical Analysis of Their Properties}",
      journal = {\apj},
         year = 2026,
        month = feb,
       volume = {998},
       number = {1},
          eid = {76},
        pages = {76},
          doi = {10.3847/1538-4357/ae2fc3},
archivePrefix = {arXiv},
       eprint = {2511.08730},
 primaryClass = {astro-ph.SR},
       adsurl = {https://ui.adsabs.harvard.edu/abs/2026ApJ...998...76C}
}

@ARTICLE{1992PhyS...46..202F,
       author = {{Feldman}, Uri},
        title = "{REVIEW:  Elemental abundances in the upper solar atmosphere}",
      journal = {\physscr},
         year = 1992,
        month = sep,
       volume = {46},
       number = {3},
        pages = {202-220},
          doi = {10.1088/0031-8949/46/3/002},
       adsurl = {https://ui.adsabs.harvard.edu/abs/1992PhyS...46..202F}
}

@ARTICLE{2009A&A...506..913H,
       author = {{Hood}, A.~W. and {Browning}, P.~K. and {van der Linden}, R.~A.~M.},
        title = "{Coronal heating by magnetic reconnection in loops with zero net current}",
      journal = {\aap},
         year = 2009,
        month = nov,
       volume = {506},
       number = {2},
        pages = {913-925},
          doi = {10.1051/0004-6361/200912285},
       adsurl = {https://ui.adsabs.harvard.edu/abs/2009A&A...506..913H}
}

@ARTICLE{1988GApFD..41..181V,
       author = {{van Ballegooijen}, A.~A.},
        title = "{Force free fields and coronal heating part I. The formation of current sheets}",
      journal = {Geophysical and Astrophysical Fluid Dynamics},
         year = 1988,
        month = jan,
       volume = {41},
       number = {3},
        pages = {181-211},
          doi = {10.1080/03091928808208850},
       adsurl = {https://ui.adsabs.harvard.edu/abs/1988GApFD..41..181V}
}

@ARTICLE{1989ApJ...338.1148M,
       author = {{Mikic}, Z. and {Schnack}, D.~D. and {van Hoven}, G.},
        title = "{Creation of Current Filaments in the Solar Corona}",
      journal = {\apj},
         year = 1989,
        month = mar,
       volume = {338},
        pages = {1148},
          doi = {10.1086/167265},
       adsurl = {https://ui.adsabs.harvard.edu/abs/1989ApJ...338.1148M}
}

@ARTICLE{2015ApJ...805...47P,
       author = {{Pontin}, D.~I. and {Hornig}, G.},
        title = "{The Structure of Current Layers and Degree of Field-line Braiding in Coronal Loops}",
      journal = {\apj},
         year = 2015,
        month = may,
       volume = {805},
       number = {1},
          eid = {47},
        pages = {47},
          doi = {10.1088/0004-637X/805/1/47},
archivePrefix = {arXiv},
       eprint = {1411.2845},
 primaryClass = {astro-ph.SR},
       adsurl = {https://ui.adsabs.harvard.edu/abs/2015ApJ...805...47P}
}

@ARTICLE{2015PASJ...67...18W,
       author = {{Welsch}, Brian T.},
        title = "{The photospheric Poynting flux and coronal heating}",
      journal = {\pasj},
         year = 2015,
        month = apr,
       volume = {67},
       number = {2},
          eid = {18},
        pages = {18},
          doi = {10.1093/pasj/psu151},
archivePrefix = {arXiv},
       eprint = {1402.4794},
 primaryClass = {astro-ph.SR},
       adsurl = {https://ui.adsabs.harvard.edu/abs/2015PASJ...67...18W}
}

@ARTICLE{2023ApJ...956L...1C,
       author = {{Chitta}, L.~P. and {Solanki}, S.~K. and {del Toro Iniesta}, J.~C. and {Woch}, J. and {Calchetti}, D. and {Gandorfer}, A. and {Hirzberger}, J. and {Kahil}, F. and {Valori}, G. and {Orozco Su{\'a}rez}, D. and {Strecker}, H. and {Appourchaux}, T. and {Volkmer}, R. and {Peter}, H. and {Mandal}, S. and {Aznar Cuadrado}, R. and {Teriaca}, L. and {Sch{\"u}hle}, U. and {Berghmans}, D. and {Verbeeck}, C. and {Zhukov}, A.~N. and {Priest}, E.~R.},
        title = "{Fleeting Small-scale Surface Magnetic Fields Build the Quiet-Sun Corona}",
      journal = {\apjl},
         year = 2023,
        month = oct,
       volume = {956},
       number = {1},
          eid = {L1},
        pages = {L1},
          doi = {10.3847/2041-8213/acf136},
archivePrefix = {arXiv},
       eprint = {2308.10982},
 primaryClass = {astro-ph.SR},
       adsurl = {https://ui.adsabs.harvard.edu/abs/2023ApJ...956L...1C}
}

@ARTICLE{2015RSPTA.37340260C,
       author = {{Cargill}, P.~J. and {Warren}, H.~P. and {Bradshaw}, S.~J.},
        title = "{Modelling nanoflares in active regions and implications for coronal heating mechanisms}",
      journal = {Philosophical Transactions of the Royal Society of London Series A},
         year = 2015,
        month = apr,
       volume = {373},
       number = {2042},
        pages = {20140260-20140260},
          doi = {10.1098/rsta.2014.0260},
       adsurl = {https://ui.adsabs.harvard.edu/abs/2015RSPTA.37340260C}
}

@ARTICLE{2015RSPTA.37340265W,
       author = {{Wilmot-Smith}, A.~L.},
        title = "{An overview of flux braiding experiments}",
      journal = {Philosophical Transactions of the Royal Society of London Series A},
         year = 2015,
        month = apr,
       volume = {373},
       number = {2042},
        pages = {20140265-20140265},
          doi = {10.1098/rsta.2014.0265},
archivePrefix = {arXiv},
       eprint = {1411.2490},
 primaryClass = {astro-ph.SR},
       adsurl = {https://ui.adsabs.harvard.edu/abs/2015RSPTA.37340265W}
}

@ARTICLE{2020LRSP...17....5P,
       author = {{Pontin}, David I. and {Hornig}, Gunnar},
        title = "{The Parker problem: existence of smooth force-free fields and coronal heating}",
      journal = {Living Reviews in Solar Physics},
         year = 2020,
        month = dec,
       volume = {17},
       number = {1},
          eid = {5},
        pages = {5},
          doi = {10.1007/s41116-020-00026-5},
       adsurl = {https://ui.adsabs.harvard.edu/abs/2020LRSP...17....5P}
}

@ARTICLE{2013ApJ...770L...1T,
       author = {{Testa}, Paola and {De Pontieu}, Bart and {Mart{\'\i}nez-Sykora}, Juan and {DeLuca}, Ed and {Hansteen}, Viggo and {Cirtain}, Jonathan and {Winebarger}, Amy and {Golub}, Leon and {Kobayashi}, Ken and {Korreck}, Kelly and {Kuzin}, Sergey and {Walsh}, Robert and {DeForest}, Craig and {Title}, Alan and {Weber}, Mark},
        title = "{Observing Coronal Nanoflares in Active Region Moss}",
      journal = {\apjl},
         year = 2013,
        month = jun,
       volume = {770},
       number = {1},
          eid = {L1},
        pages = {L1},
          doi = {10.1088/2041-8205/770/1/L1},
archivePrefix = {arXiv},
       eprint = {1305.1687},
 primaryClass = {astro-ph.SR},
       adsurl = {https://ui.adsabs.harvard.edu/abs/2013ApJ...770L...1T}
}

@ARTICLE{1996JGR...10113445G,
       author = {{Galsgaard}, Klaus and {Nordlund}, {\r{A}}ke},
        title = "{Heating and activity of the solar corona 1. Boundary shearing of an initially homogeneous magnetic field}",
      journal = {\jgr},
         year = 1996,
        month = jun,
       volume = {101},
       number = {A6},
        pages = {13445-13460},
          doi = {10.1029/96JA00428},
       adsurl = {https://ui.adsabs.harvard.edu/abs/1996JGR...10113445G}
}

@ARTICLE{2012ApJ...747..109N,
       author = {{Ng}, C.~S. and {Lin}, L. and {Bhattacharjee}, A.},
        title = "{High-Lundquist Number Scaling in Three-dimensional Simulations of Parker's Model of Coronal Heating}",
      journal = {\apj},
         year = 2012,
        month = mar,
       volume = {747},
       number = {2},
          eid = {109},
        pages = {109},
          doi = {10.1088/0004-637X/747/2/109},
archivePrefix = {arXiv},
       eprint = {1106.0515},
 primaryClass = {astro-ph.SR},
       adsurl = {https://ui.adsabs.harvard.edu/abs/2012ApJ...747..109N}
}

@ARTICLE{2008ApJ...677.1348R,
       author = {{Rappazzo}, A.~F. and {Velli}, M. and {Einaudi}, G. and {Dahlburg}, R.~B.},
        title = "{Nonlinear Dynamics of the Parker Scenario for Coronal Heating}",
      journal = {\apj},
         year = 2008,
        month = apr,
       volume = {677},
       number = {2},
        pages = {1348-1366},
          doi = {10.1086/528786},
archivePrefix = {arXiv},
       eprint = {0709.3687},
 primaryClass = {astro-ph},
       adsurl = {https://ui.adsabs.harvard.edu/abs/2008ApJ...677.1348R}
}

@ARTICLE{2007ApJ...657L..47R,
       author = {{Rappazzo}, A.~F. and {Velli}, M. and {Einaudi}, G. and {Dahlburg}, R.~B.},
        title = "{Coronal Heating, Weak MHD Turbulence, and Scaling Laws}",
      journal = {\apjl},
         year = 2007,
        month = mar,
       volume = {657},
       number = {1},
        pages = {L47-L51},
          doi = {10.1086/512975},
archivePrefix = {arXiv},
       eprint = {astro-ph/0701872},
 primaryClass = {astro-ph},
       adsurl = {https://ui.adsabs.harvard.edu/abs/2007ApJ...657L..47R}
}

@ARTICLE{2010A&A...516A...5W,
       author = {{Wilmot-Smith}, A.~L. and {Pontin}, D.~I. and {Hornig}, G.},
        title = "{Dynamics of braided coronal loops. I. Onset of magnetic reconnection}",
      journal = {\aap},
         year = 2010,
        month = jun,
       volume = {516},
          eid = {A5},
        pages = {A5},
          doi = {10.1051/0004-6361/201014041},
archivePrefix = {arXiv},
       eprint = {1001.1717},
 primaryClass = {astro-ph.SR},
       adsurl = {https://ui.adsabs.harvard.edu/abs/2010A&A...516A...5W}
}

@ARTICLE{2011A&A...525A..57P,
       author = {{Pontin}, D.~I. and {Wilmot-Smith}, A.~L. and {Hornig}, G. and {Galsgaard}, K.},
        title = "{Dynamics of braided coronal loops. II. Cascade to multiple small-scale reconnection events}",
      journal = {\aap},
         year = 2011,
        month = jan,
       volume = {525},
          eid = {A57},
        pages = {A57},
          doi = {10.1051/0004-6361/201014544},
archivePrefix = {arXiv},
       eprint = {1003.5784},
 primaryClass = {astro-ph.SR},
       adsurl = {https://ui.adsabs.harvard.edu/abs/2011A&A...525A..57P}
}

@ARTICLE{2015ApJ...808..134C,
       author = {{Candelaresi}, S. and {Pontin}, D.~I. and {Hornig}, G.},
        title = "{Magnetic Field Relaxation and Current Sheets in an Ideal Plasma}",
      journal = {\apj},
         year = 2015,
        month = aug,
       volume = {808},
       number = {2},
          eid = {134},
        pages = {134},
          doi = {10.1088/0004-637X/808/2/134},
archivePrefix = {arXiv},
       eprint = {1505.03043},
 primaryClass = {astro-ph.SR},
       adsurl = {https://ui.adsabs.harvard.edu/abs/2015ApJ...808..134C}
}

@ARTICLE{2005ApJ...618.1020G,
       author = {{Gudiksen}, Boris Vilhelm and {Nordlund}, {\r{A}}ke},
        title = "{An Ab Initio Approach to the Solar Coronal Heating Problem}",
      journal = {\apj},
         year = 2005,
        month = jan,
       volume = {618},
       number = {2},
        pages = {1020-1030},
          doi = {10.1086/426063},
archivePrefix = {arXiv},
       eprint = {astro-ph/0407266},
 primaryClass = {astro-ph},
       adsurl = {https://ui.adsabs.harvard.edu/abs/2005ApJ...618.1020G}
}

@ARTICLE{2005ApJ...618.1031G,
       author = {{Gudiksen}, Boris Vilhelm and {Nordlund}, {\r{A}}ke},
        title = "{An AB Initio Approach to Solar Coronal Loops}",
      journal = {\apj},
         year = 2005,
        month = jan,
       volume = {618},
       number = {2},
        pages = {1031-1038},
          doi = {10.1086/426064},
archivePrefix = {arXiv},
       eprint = {astro-ph/0407267},
 primaryClass = {astro-ph},
       adsurl = {https://ui.adsabs.harvard.edu/abs/2005ApJ...618.1031G}
}

@ARTICLE{2011A&A...530A.112B,
       author = {{Bingert}, S. and {Peter}, H.},
        title = "{Intermittent heating in the solar corona employing a 3D MHD model}",
      journal = {\aap},
         year = 2011,
        month = jun,
       volume = {530},
          eid = {A112},
        pages = {A112},
          doi = {10.1051/0004-6361/201016019},
archivePrefix = {arXiv},
       eprint = {1103.6042},
 primaryClass = {physics.space-ph},
       adsurl = {https://ui.adsabs.harvard.edu/abs/2011A&A...530A.112B}
}

@ARTICLE{2015ApJ...811..106H,
       author = {{Hansteen}, V. and {Guerreiro}, N. and {De Pontieu}, B. and {Carlsson}, M.},
        title = "{Numerical Simulations of Coronal Heating through Footpoint Braiding}",
      journal = {\apj},
         year = 2015,
        month = oct,
       volume = {811},
       number = {2},
          eid = {106},
        pages = {106},
          doi = {10.1088/0004-637X/811/2/106},
archivePrefix = {arXiv},
       eprint = {1508.07234},
 primaryClass = {astro-ph.SR},
       adsurl = {https://ui.adsabs.harvard.edu/abs/2015ApJ...811..106H}
}

@ARTICLE{2017ApJ...834...10R,
       author = {{Rempel}, M.},
        title = "{Extension of the MURaM Radiative MHD Code for Coronal Simulations}",
      journal = {\apj},
         year = 2017,
        month = jan,
       volume = {834},
       number = {1},
          eid = {10},
        pages = {10},
          doi = {10.3847/1538-4357/834/1/10},
archivePrefix = {arXiv},
       eprint = {1609.09818},
 primaryClass = {astro-ph.SR},
       adsurl = {https://ui.adsabs.harvard.edu/abs/2017ApJ...834...10R}
}

@ARTICLE{2026ApJ..1004..149C,
       author = {{Chen}, Feng},
        title = "{Data-driven Radiative Magnetohydrodynamics Simulations with the MURaM Code: Coronal Heating and Dynamics in an Emerging Active Region}",
      journal = {\apj},
         year = 2026,
        month = jun,
       volume = {1004},
       number = {2},
          eid = {149},
        pages = {149},
          doi = {10.3847/1538-4357/ae6daa},
archivePrefix = {arXiv},
       eprint = {2511.02362},
 primaryClass = {astro-ph.SR},
       adsurl = {https://ui.adsabs.harvard.edu/abs/2026ApJ..1004..149C}
}

@ARTICLE{2019A&A...624L..12W,
       author = {{Warnecke}, J. and {Peter}, H.},
        title = "{Data-driven model of the solar corona above an active region}",
      journal = {\aap},
         year = 2019,
        month = apr,
       volume = {624},
          eid = {L12},
        pages = {L12},
          doi = {10.1051/0004-6361/201935385},
archivePrefix = {arXiv},
       eprint = {1903.00455},
 primaryClass = {astro-ph.SR},
       adsurl = {https://ui.adsabs.harvard.edu/abs/2019A&A...624L..12W}
}

@ARTICLE{2026MNRAS.545f2180B,
       author = {{Breu}, C.~A. and {Pontin}, D.~I. and {Priest}, E. and {De Moortel}, I.},
        title = "{On the complex nature of coronal heating}",
      journal = {\mnras},
         year = 2026,
        month = jan,
       volume = {545},
       number = {3},
          eid = {staf2180},
        pages = {staf2180},
          doi = {10.1093/mnras/staf2180},
archivePrefix = {arXiv},
       eprint = {2512.17880},
 primaryClass = {astro-ph.SR},
       adsurl = {https://ui.adsabs.harvard.edu/abs/2026MNRAS.545f2180B}
}

@ARTICLE{2025ApJ...994..139J,
       author = {{Johnston}, Craig D. and {Daldorff}, Lars K.~S. and {Klimchuk}, James A. and {Mondal}, Shanwlee Sow and {Barnes}, Will T. and {Leake}, James E. and {Reid}, Jack and {Parker}, Jacob D.},
        title = "{Self-Consistent Heating of the Magnetically Closed Solar Corona: Generation of Nanoflares, Thermodynamic Response of the Plasma and Observational Signatures}",
      journal = {\apj},
         year = 2025,
        month = dec,
       volume = {994},
       number = {2},
          eid = {139},
        pages = {139},
          doi = {10.3847/1538-4357/ae08a2},
archivePrefix = {arXiv},
       eprint = {2508.12952},
 primaryClass = {astro-ph.SR},
       adsurl = {https://ui.adsabs.harvard.edu/abs/2025ApJ...994..139J}
}

\end{document}